\documentclass[a4paper,11pt]{article}
\usepackage{jcappub} % for details on the use of the package, please see the JINST-author-manual
\usepackage{lineno}
\usepackage{gensymb}
\usepackage{amsmath}
\title{\boldmath Two-component $\gamma$-Ray Structure from the CR Sources}

\author[b]{Lin Nie}
\author[b,d]{Ming-Jun Chen}
\author[b,d]{Zong-Kang Zeng}
\author[b,c,d]{Yi-Qing Guo}

\affiliation[b]{Key Laboratory of Particle Astrophysics,\\
Institute of High Energy Physics, Chinese Academy of Sciences,\\
Beijing, 100049, China}
\affiliation[c]{University of Chinese Academy of Sciences,\\ 
Beijing 100049, China }
\affiliation[d]{TIANFU Cosmic Ray Research Center,\\
Chengdu 610000, China}
\emailAdd{mjchen@ihep}
\emailAdd{zengzk@ihep.ac.cn}
\abstract{With the advent of a new generation of observational experiments and the accumulation of data, an increasing number of cosmic-ray sources have been found to exhibit pronounced two-component structures in their gamma-ray spectra above GeV energies. To investigate whether these structures share a common physical origin, we conduct a statistical study of sources exhibiting such features using a two-component model of cosmic-ray sources and examine their gamma-ray emission mechanisms. We find that the ambient environment significantly affects the spectral morphology. Sources located in low-density environments but surrounded by dense molecular clumps exhibit two clearly separated spectral peaks. In contrast, sources embedded in dense environments with additional dense molecular clumps nearby retain two-component spectral features, but the components overlap more smoothly, with the high-energy component appearing primarily as a relatively broad shoulder rather than a distinct peak. Our results lend further support to a physical picture in which high-energy cosmic rays are predominantly supplied by local sources, whereas low-energy cosmic rays arise mainly from the collective contribution of sources in other regions of the Galaxy. We also calculate the neutrino spectra of these sources, providing theoretical predictions for future tests of the model through observations of point-source and diffuse neutrino emission.
}


\begin{document}
\maketitle
\flushbottom

\section{Introduction}
\label{sec:intro}
In recent years, increasingly sensitive cosmic-ray and gamma-ray experiments, particularly space-based experiments such as AMS-02, Fermi-LAT, PAMELA, ATIC, CREAM, and China’s Dark Matter Particle Explorer (DAMPE)\citep{2015PhRvL.114q1103A,2015PhRvL.115u1101A,2017PhRvL.119y1101A,2018PhRvL.120b1101A,2018PhRvL.121e1103A,2020PhRvL.124u1102A,2021PhRvL.126d1104A}, have enabled progressively more precise measurements of cosmic-ray and gamma-ray spectra. These measurements have revealed a series of “anomalies” that are difficult to explain within conventional cosmic-ray propagation models. In particular, the spectra of different cosmic-ray components deviate from simple power laws above a few hundred GeV, exhibiting pronounced spectral hardening that suggests a possible two-component structure. Similar features have also been observed in the gamma-ray spectra of an increasing number of point sources \citep{2025ApJ...982L..33C,2023Univ....9...98S,2021MNRAS.502..472A,2025ApJ...988...93N}, providing new clues to the origin and propagation of cosmic rays.

To provide a unified explanation for these phenomena, spatially dependent cosmic-ray propagation models have been developed \citep{2018PhRvD..97f3008G,2024PhRvD.109f3001Y,2024ApJ...974..276N,2023ApJ...956...75Q}. These models attribute the spectral hardening of high-energy cosmic rays to spatial inhomogeneities in their propagation and predict that two-component spectra may be widespread throughout the Galaxy. The underlying physical picture is that the halo surrounding a cosmic-ray source plays a dual role: it acts as a “barrier” that inhibits the entry of external cosmic rays and as a “potential well” that confines particles produced locally or entering the halo. Cosmic rays from distant regions propagate primarily through the outer Galactic halo before entering and becoming confined within local source halos. Consequently, cosmic rays in a local halo comprise two components: a cosmic-ray “sea,” supplied collectively by numerous Galactic sources and dominant at low energies, and a component supplied by the local source that dominates at high energies. Together, these components produce a two-component spectrum, and the resulting gamma-ray emission is expected to exhibit similar features.

This framework provides a self-consistent explanation for the spectral hardening of different cosmic-ray components and further suggests that similar two-component structures may extend to the scales of individual sources and their surrounding halos. Observations of gamma-ray emission from individual cosmic-ray sources and their neighboring regions therefore offer a means of testing this physical picture.

Recently, the LHAASO Collaboration reported observations of several sources located in dense environments, some of which, including the supernova remnant Cas A, exhibit pronounced two-component spectral structures. To explain these observations, we proposed a two-component emission model in our previous work and obtained a self-consistent description of the spectral features of Cas A, Tycho, and IC 443 \citep{2026JCAP...10..022N} (hereafter Paper I). In this model, the high-energy emission from a source consists primarily of two contributions. The first arises from particles injected from the thermal population and subsequently accelerated at the source, which produce gamma rays through inverse Compton scattering or proton–proton (p–p) interactions. The second originates from the Galactic cosmic-ray background: background protons are reaccelerated by local shocks or bow shocks and subsequently interact with the surrounding gas to produce gamma rays. The former contribution dominates the high-energy component of the spectrum, whereas the latter contributes primarily to the low-energy component.

In this work, we apply this two-component model to several sources recently observed by LHAASO and other experiments to explain their two-component spectral features and further assess the model’s applicability. We also investigate how widely two-power-law spectral structures occur in Galactic cosmic rays across different spatial scales. In addition, we calculate the neutrino spectra of these sources, providing predictions that can be tested by future observations of point-source neutrino emission and used to further constrain the model through the neutrino messenger. This paper is organized as follows. Section \ref{sec:method} describes our methods. Section \ref{sec:result} presents and discusses the results. Section \ref{sec:conclusion} summarizes our findings and outlines prospects for future research.

\section{MODEL \& METHODOLOGY} \label{sec:method}

\subsection{Physical Picture and Basic Assumptions}

Within the two-component framework proposed in Paper I \citep{2026JCAP...10..022N}, we further investigate the contribution of Galactic background cosmic rays to the nonthermal emission of cosmic-ray sources in dense-cloud environments. Compared with Paper I, the present study includes six additional sources: $\gamma$ Cygni, CTB 80, J1729, J852, J857, and J1912. We also calculate the neutrino spectra produced by inelastic proton–proton ($p$–$p$) interactions in these sources to establish a physical connection between their gamma-ray and neutrino emission.

The model distinguishes two populations of nonthermal particles with different origins: Galactic background cosmic rays already present in the surrounding dense gas before the arrival of the shock, and particles injected and accelerated by the source itself. For the former population, shock reacceleration and adiabatic compression can increase both particle energies and number densities, thereby enhancing gamma-ray emission from neutral-pion decay and neutrino emission from charged-meson decay following interactions with dense gas. For the latter population, high-energy electrons accelerated in low-density regions can contribute to the gamma-ray emission through inverse Compton scattering. The hadronic gamma-ray and neutrino components are calculated using the same proton spectra and gas parameters, without introducing an independent normalization for the neutrino flux.

For sources to which the shock–cloud interaction picture applies, the ambient medium is approximated by two regions. The low-density region provides an environment for faster shock propagation and fresh particle acceleration, whereas the dense-cloud region hosts the cloud shock, downstream radiative cooling, and subsequent compression. Each region has its own gas density, magnetic field, and effective emitting volume. The acceleration environment and associated astrophysical object of each source are determined by the corresponding observational constraints; the six additional sources are not required to share identical dynamical histories. The differential gamma-ray flux in the baseline model is

\begin{equation}
\frac{d\Phi_\gamma}{dE_\gamma}
=
\frac{d\Phi_{\pi^0,\mathrm{re}}}{dE_\gamma}
+
\frac{d\Phi_{\mathrm{IC},\mathrm{fresh}}}{dE_\gamma}.
\label{eq1}
\end{equation}

Here, the subscript $\mathrm{re}$ denotes reaccelerated and compressed background protons, while $\mathrm{fresh}$ denotes freshly accelerated electrons. The relative contributions of the two components are jointly determined by the particle spectra and local environments; they are not assumed a priori to dominate the same photon-energy ranges in every source. The term “two-component” refers here to the distinction between particle origins. It does not require the total gamma-ray spectrum to exhibit two clearly separated peaks, nor does it require each observed spatial component to correspond to a distinct particle population.

\subsection{Galactic Background Cosmic Rays}

The background cosmic-ray spectra are calculated using the same spatially dependent propagation model as in Paper I. Galactic-scale transport includes spatial diffusion, reacceleration in momentum space, and the relevant particle losses. GALPROP is used to obtain the interstellar proton spectrum at each source location \citep{1998ApJ...509..212S}. The spatial diffusion coefficient is written as \citep{2018PhRvD..97f3008G}

\begin{equation}
D_{xx}(r,z,\mathcal{R})
=
D_0 F(r,z)\,\beta^\eta
\left(\frac{\mathcal{R}}{\mathcal{R}_0}\right)^{\delta_0 F(r,z)},
\label{eq2}
\end{equation}

where $r$ and $z$ are Galactocentric cylindrical coordinates, $\mathcal{R}$ is rigidity, $\beta=v/c$, and $\mathcal{R}_0$ is the reference rigidity. The spatial modulation function is

\begin{equation}
\begin{aligned}
F(r,z)
&=
\begin{cases}
g(r,z)+[1-g(r,z)]
\left(\dfrac{|z|}{\xi z_0}\right)^n,
& |z|\leq \xi z_0,\\[6pt]
1,
& |z|>\xi z_0,
\end{cases}\\
g(r,z)
&=
\frac{N_m}{1+f_{\mathrm{src}}(r,z)}.
\end{aligned}
\label{eq3}
\end{equation}

Here, $f_{\mathrm{src}}$ is the spatial source distribution normalized according to the convention adopted in Paper I, $z_0$ is the propagation-halo half-height, and $\xi z_0$ is the inner-halo half-thickness. The parameters $N_m$ and $n$ control the normalization of the inner-halo diffusion coefficient and the smoothness of its transition to the outer halo, respectively. For consistency with Paper I, we adopt the baseline propagation parameters listed in its Table 1: $D_0=4.8\times10^{28}\,\mathrm{cm^2\,s^{-1}}$, $\delta_0=0.63$, $N_m=0.24$, $\xi=0.1$, $n=4$, $v_A=8\,\mathrm{km\,s^{-1}}$, and $z_0=4.5\,\mathrm{kpc}$. The reference rigidity, low-energy velocity correction, and source-injection settings are kept consistent with the numerical configuration used in Paper I.

Let $\psi_p(\mathbf{r},p)=dn_p/dp$ denote the proton number density per unit momentum. The isotropic phase-space distribution at the source position $\mathbf{r}_s$ is then

\begin{equation}
f_{\mathrm{GCR}}(p;\mathbf{r}_s)
=
\frac{\psi_p(\mathbf{r}_s,p)}{4\pi p^2}.
\label{eq4}
\end{equation}

This spectrum serves as the upstream seed-particle distribution for the cloud shock. We use the interstellar spectrum here. Solar modulation is applied only when comparing the propagation model with near-Earth observations and is not used to modify the upstream spectrum at the source location. The background spectrum and the particle populations explicitly attributed to the source under study must be treated consistently to avoid double counting particles produced by the same source.

\subsection{Shock Reacceleration and Adiabatic Compression of Background Protons}

In the local planar-shock approximation, the shock is located at $x=0$, with the $x$-axis directed along the shock normal. Neglecting particle losses within the acceleration layer, the steady-state transport equation is

\begin{equation}
u\frac{\partial f}{\partial x}
=
\frac{\partial}{\partial x}
\left(
\kappa\frac{\partial f}{\partial x}
\right)
+
\frac{1}{3}\frac{\partial u}{\partial x}\,
p\frac{\partial f}{\partial p}.
\label{eq5}
\end{equation}

Here, $u$ is the gas velocity in the shock rest frame, and $\kappa$ is the particle diffusion coefficient near the shock. The latter has a different physical meaning from the Galactic-scale diffusion coefficient in Equation (2). Let $u_1$ and $u_2$ denote the upstream and downstream velocities, respectively, and define the shock compression ratio as $r_{\mathrm{sh}}=u_1/u_2$. Adopting the low-momentum boundary convention of Paper I, the distribution of reaccelerated particles at the shock is \citep{1978ApJ...221L..29B}

\begin{equation}
f_{\mathrm{acc}}(p)
=
q\,p^{-q}
\int_{p_{\min}}^p
dp'\,p'^{\,q-1}f_{\mathrm{GCR}}(p'),
\qquad
q=\frac{3r_{\mathrm{sh}}}{r_{\mathrm{sh}}-1},
\label{eq6}
\end{equation}

where $p_{\min}$ is the minimum momentum included in the reacceleration calculation. This expression describes the spectral reshaping of background particles by the shock, with the degree of hardening determined by the seed-spectrum slope and shock compression ratio. In particular, if the seed phase-space distribution is steeper than $p^{-q}$, fully developed steady-state reacceleration drives the high-momentum spectrum toward $p^{-q}$. A low shock velocity alone therefore does not justify assuming that the reaccelerated spectrum retains the original background spectral shape.

For cloud shocks with a significant neutral component, where ion–neutral damping is applicable, we adopt the effective spectral-steepening prescription \citep{2019MNRAS.482.3843T,2011NatCo...2..194M}:

\begin{equation}
f_{\mathrm{sh}}(p)
=
f_{\mathrm{acc}}(p)\,\mathcal{B}(p),
\qquad
\mathcal{B}(p)
=
\begin{cases}
1, & p\leq p_{\mathrm{br}},\\[3pt]
p_{\mathrm{br}}/p, & p>p_{\mathrm{br}}.
\end{cases}
\label{eq7}
\end{equation}

Here, $p_{\mathrm{br}}$ is the momentum marking the damping-related spectral break, and the limit $p_{\mathrm{br}}\rightarrow\infty$ corresponds to the absence of this steepening. This feature is distinct from the electron spectral cutoff introduced below and is not assumed to occur in every dense cloud.

Radiative cooling further compresses the downstream gas. Let $n_0$ be the upstream hydrogen-nucleus density in the cloud, $n_d=r_{\mathrm{sh}}n_0$ the density immediately downstream of the shock, and $n_m$ the density in the final cooled and compressed layer. The additional compression factor is 

\begin{equation}
s
\equiv
\frac{n_m}{n_d}
=
\frac{n_m}{r_{\mathrm{sh}}n_0}.
\label{eq}
\end{equation}

Provided that particles remain confined to the gas, their momentum distribution is approximately isotropic, and losses during compression are negligible, their momenta scale as $p\rightarrow s^{1/3}p$. Defining $n_{p,\mathrm{sh}}(p)=4\pi p^2f_{\mathrm{sh}}(p)$, the compressed proton spectrum per unit volume and momentum is \citep{1982ApJ...260..625B}

\begin{equation}
n_{p,\mathrm{ad}}(p)
=
s^{2/3}
n_{p,\mathrm{sh}}\!\left(s^{-1/3}p\right).
\label{eq9}
\end{equation}

\subsection{Proton Accumulation and Energy Losses in the Compressed Region}

Let $R_s$ be the effective radius of the source region, with a corresponding reference volume $V_s=4\pi R_s^3/3$. For supernova remnants with a well-defined shell, $R_s$ is taken to be the remnant radius. We use $\phi V_s$ to denote the pre-compression volume occupied by the cloud gas processed over the interaction time $t_c$, where $\phi$ is the effective volume fraction. Mass conservation requires the final volume of the compressed region to be

\begin{equation}
V_m
=
\frac{n_0}{n_m}\,\phi V_s.
\label{eq10}
\end{equation}

Assuming fixed cloud-shock parameters and approximating the entry of cloud material into the compressed region by a constant effective processing rate, the proton input rate is

\begin{equation}
Q_p(p)
=
\frac{V_m}{t_c}\,n_{p,\mathrm{ad}}(p)
=
\frac{n_0\phi V_s}{n_m t_c}\,
n_{p,\mathrm{ad}}(p).
\label{11}
\end{equation}

Define $N_p(p,t)=dN_p/dp$ as the volume-integrated proton momentum spectrum in the compressed region. Its evolution obeys

\begin{equation}
\frac{\partial N_p}{\partial t}
=
\frac{\partial}{\partial p}
\left[
b_p(p)N_p(p,t)
\right]
+
Q_p(p),
\qquad
b_p(p)\equiv-\frac{dp}{dt}>0.
\label{12}
\end{equation}

The loss term includes ionization, Coulomb interactions, and inelastic nuclear collisions. Inelastic collisions are treated using the continuous-loss approximation in this equation. If the loss rate is expressed in terms of proton kinetic energy as $b_T=-dT_p/dt$, it must be converted to momentum space using $b_p=b_T/v_p$. Within this approximation, no additional catastrophic-loss term describing the same collisions is included, thereby avoiding double counting.

The initial condition $N_p(p,0)=0$ represents the absence of accumulated processed cloud material, and the emission is evaluated at $t=t_c$. In the loss-free limit, $N_p(p,t_c)=V_m n_{p,\mathrm{ad}}(p)$, which provides a normalization check. Equation \ref{12} treats compression and subsequent losses as separate stages and neglects escape from the compressed region. The compression timescale must therefore be checked against the loss timescale, and $t_c$ must be compared with the particle escape time. Here, $t_c$ denotes the duration of the interaction between the cloud and the source and generally differs from the age of the source itself, such as a supernova remnant.

\subsection{Freshly Accelerated Particles}

The present-day, volume-integrated energy spectra of freshly accelerated particles are described by a smoothly broken power law with an exponential cutoff:

\begin{equation}
\begin{aligned}
N_i(E_i)
={}&
K_i
\left(\frac{E_i}{E_0}\right)^{-\alpha_1}
\left[
1+\left(\frac{E_i}{E_{b,i}}\right)^2
\right]^{(\alpha_1-\alpha_2)/2}
\\
&\times
\exp\!\left(-\frac{E_i}{E_{\mathrm{cut},i}}\right),
\end{aligned}
\label{eq13}
\end{equation}

where $N_i=dN_i/dE_i$, and $i$ denotes protons ($p$) or electrons ($e$). Here, $E_i$ is the particle energy, $E_0=1\,\mathrm{TeV}$ is the reference energy, and $K_i$ is the spectral normalization. The parameters $\alpha_1$ and $\alpha_2$ are the spectral indices below and above the break, respectively, while $E_{b,i}$ and $E_{\mathrm{cut},i}$ are the break and cutoff energies.

Equation \ref{eq13} represents an effective particle spectrum after acceleration, cooling, and transport. It is not directly identified with the injection rate of thermal particles, nor is its cutoff assumed a priori to originate from cooling. Electrons produce inverse Compton emission by scattering cosmic microwave background, infrared, and optical photons. Klein–Nishina effects are therefore included in the calculation.

\subsection{Nonthermal Emission Calculations}

For a compressed layer with uniform density $n_m$, the gamma-ray flux from neutral-pion decay produced by proton interactions is

\begin{equation}
\frac{d\Phi_{\pi^0,\mathrm{re}}}{dE_\gamma}
=
\frac{n_m}{4\pi d^2}
\int_{T_{\mathrm{th}}}^{\infty}
dT_p\,
v_p(T_p)\,
N_{p,T}(T_p)\,
\frac{d\sigma_{pp\rightarrow\gamma}(T_p,E_\gamma)}
     {dE_\gamma},
\label{eq14}
\end{equation}

where $d$ is the source distance, $T_{\mathrm{th}}$ is the threshold kinetic energy for neutral-pion production, and $N_{p,T}=dN_p/dT_p=N_p/v_p$. We adopt a cross-section parameterization covering energies from threshold to the high-energy regime.

The electron inverse Compton flux is

\begin{equation}
\frac{d\Phi_{\mathrm{IC}}}{dE_\gamma}
=
\frac{c}{4\pi d^2}
\int dE_e\,N_e(E_e)
\int d\epsilon\,n_{\mathrm{ph}}(\epsilon)
\frac{d\sigma_{\mathrm{IC}}(E_e,\epsilon,E_\gamma)}
     {dE_\gamma},
\label{eq15}
\end{equation}

where $n_{\mathrm{ph}}$ is the target-photon number density per unit photon energy. The scattering kernel assumes isotropic electron and photon distributions. Gamma-ray emission is calculated using Naima \citep{2015ICRC...34..922Z} or an equivalent implementation. All gamma-ray spectra are expressed as $E_\gamma^2\,d\Phi_\gamma/dE_\gamma$ for comparison with the observed spectral energy distributions.

\subsection{ Neutrino Spectra from $p$–$p$ Interactions}

In addition to producing gamma rays through neutral-pion decay, the same inelastic $p$–$p$ collisions also produce charged mesons. Neutrinos arise primarily from charged pions and the muons produced in their decays. The principal decay chains are

\begin{equation}
\begin{aligned}
p+p
&\rightarrow
\pi^0+\pi^\pm+X,
\\
\pi^+
&\rightarrow
\mu^++\nu_\mu
\rightarrow
e^++\nu_e+\bar{\nu}_\mu+\nu_\mu,
\\
\pi^-
&\rightarrow
\mu^-+\bar{\nu}_\mu
\rightarrow
e^-+\bar{\nu}_e+\nu_\mu+\bar{\nu}_\mu.
\end{aligned}
\label{eq16}
\end{equation}

Here, $X$ denotes other hadronic final-state particles, and the first line schematically represents the different meson-production channels. We calculate neutrino emission using the high-energy parameterization of \cite{2006PhRvC..74c4318O}, treating the muon-decay and direct charged-pion-decay channels separately. Let $N_{p,E,j}(E_p)=dN_{p,j}/dE_p$ be the volume-integrated energy spectrum of interacting protons in source $j$, $n_{m,j}$ the target-gas hydrogen-nucleus density, and $d_j$ the source distance. Adopting the high-energy approximation $v_p\simeq c$, the differential flux from each channel, including geometric dilution but before flavor oscillations, is

\begin{equation}
\begin{aligned}
\mathcal{I}_{h,j}(E_\nu)
={}&
\frac{c\,n_{m,j}}{4\pi d_j^2}
\int_{E_{p,\min}^{(h)}}^{E_{p,\mathrm{up}}}
\frac{dE_p}{E_p}\,
N_{p,E,j}(E_p)
\\
&\times
\sigma_{\mathrm{inel}}(E_p)
F_h\!\left(\frac{E_\nu}{E_p},E_p\right),
\qquad
h=\mu d,\pi d.
\end{aligned}
\label{eq17}
\end{equation}

The muon-decay channel is evaluated over $0<x<1$, with the lower integration limit $E_{p,\min}^{(\mu d)}=\max(E_\nu,0.1\,\mathrm{TeV})$. The direct pion-decay channel requires $0<x<0.427$, giving $E_{p,\min}^{(\pi d)}=\max(E_\nu/0.427,0.1\,\mathrm{TeV})$. If the lower limit is greater than or equal to the upper limit, the corresponding channel is set to zero. The inelastic cross section is $\sigma_{\mathrm{inel}}(E_p)=(34.3+1.88L+0.25L^2)\times10^{-27}\,\mathrm{cm^2}$, where $L=\ln(E_p/1\,\mathrm{TeV})$. The functions $F_{\mu d}(x,E_p)$ and $F_{\pi d}(x,E_p)$ denote the muon-decay and direct pion-decay yield kernels, respectively; the former is also used to approximate the electron-flavor neutrino yield. Here, $x=E_\nu/E_p$. Both kernels include the corresponding neutrino and antineutrino contributions, so no additional factor of two is applied to account for the two charge states.

Within this pion–muon decay approximation, we define $\mathcal{F}_{\alpha,j}^{(0)}$ as the flux corresponding to the source flavor composition after geometric dilution by $1/(4\pi d_j^2)$, rather than the intrinsic production rate at the source. The electron-, muon-, and tau-flavor fluxes before oscillations satisfy

\begin{equation}
\begin{aligned}
\mathcal{F}_{e,j}^{(0)}(E_\nu)
&=
\mathcal{I}_{\mu d,j}(E_\nu),
\\
\mathcal{F}_{\mu,j}^{(0)}(E_\nu)
&=
\mathcal{I}_{\mu d,j}(E_\nu)
+
\mathcal{I}_{\pi d,j}(E_\nu),
\\
\mathcal{F}_{\tau,j}^{(0)}(E_\nu)
&=
0.
\end{aligned}
\label{eq17}
\end{equation}

The all-flavor flux is obtained by summing the electron- and muon-flavor contributions before oscillations. After propagation, we adopt approximate flavor equipartition:

\begin{equation}
\begin{aligned}
\mathcal{F}_{\mathrm{all},j}
&=
\mathcal{F}_{e,j}^{(0)}
+
\mathcal{F}_{\mu,j}^{(0)}
=
2\mathcal{I}_{\mu d,j}
+
\mathcal{I}_{\pi d,j},
\\
\mathcal{F}_{\alpha,j}^{\oplus,\mathrm{approx}}
&=
\frac{1}{3}\mathcal{F}_{\mathrm{all},j},
\qquad
\alpha=e,\mu,\tau.
\end{aligned}
\label{eq18}
\end{equation}

The superscript $\oplus,\mathrm{approx}$ explicitly denotes the flavor-equipartition approximation for the flux at Earth. Only the pion–muon decay channels described above are included. Contributions from kaons and other secondary particles are neglected, and pion or muon cooling, absorption during propagation, Earth absorption, and detector response are not explicitly calculated.

\section{Results and Discussion} \label{sec:result}
With the baseline Galactic propagation model fixed, the source-region parameters to be constrained include the compression factor $s$, the effective cloud volume fraction $\phi$, and the spectral parameters of freshly accelerated electrons. For sources whose high-energy component is interpreted as hadronic emission, the spectral parameters of freshly accelerated protons must also be constrained. Other parameters, including the upstream cloud density, shock compression ratio, damping-related break momentum, interaction time, source distance, and radius, are preferentially determined from independent observations or explicit physical assumptions. Their uncertainties are assessed where necessary to evaluate their impact on the results. We perform Bayesian parameter inference using the Markov chain Monte Carlo (MCMC) method, employing the affine-invariant ensemble sampler implemented in the open-source Python package emcee \citep{2013PASP..125..306F}. Assuming Gaussian errors on the observed flux points, we construct the log-likelihood function as follows:
\begin{equation}
\ln \mathcal{L}(\boldsymbol{\theta})
=
-\frac{1}{2}\sum_i
\left[
\frac{\left(y_i-m_i(\boldsymbol{\theta})\right)^2}{\sigma_i^2}
+
\ln\left(2\pi\sigma_i^2\right)
\right].
\end{equation}
The observed and model-predicted values in this expression use the same flux definition. The MCMC sampling uses $N_{\mathrm{walkers}}=20$ walkers, each evolved for 20,000 steps. The first 1,000 steps are discarded as burn-in to reduce the influence of the initial conditions. Parameter uncertainties are bounded by the 16th and 84th percentiles of the posterior distributions, corresponding to approximately $68\%$ credible intervals.
\begin{figure}[t]%[htbp]
\centering
\includegraphics[width=0.98\linewidth]{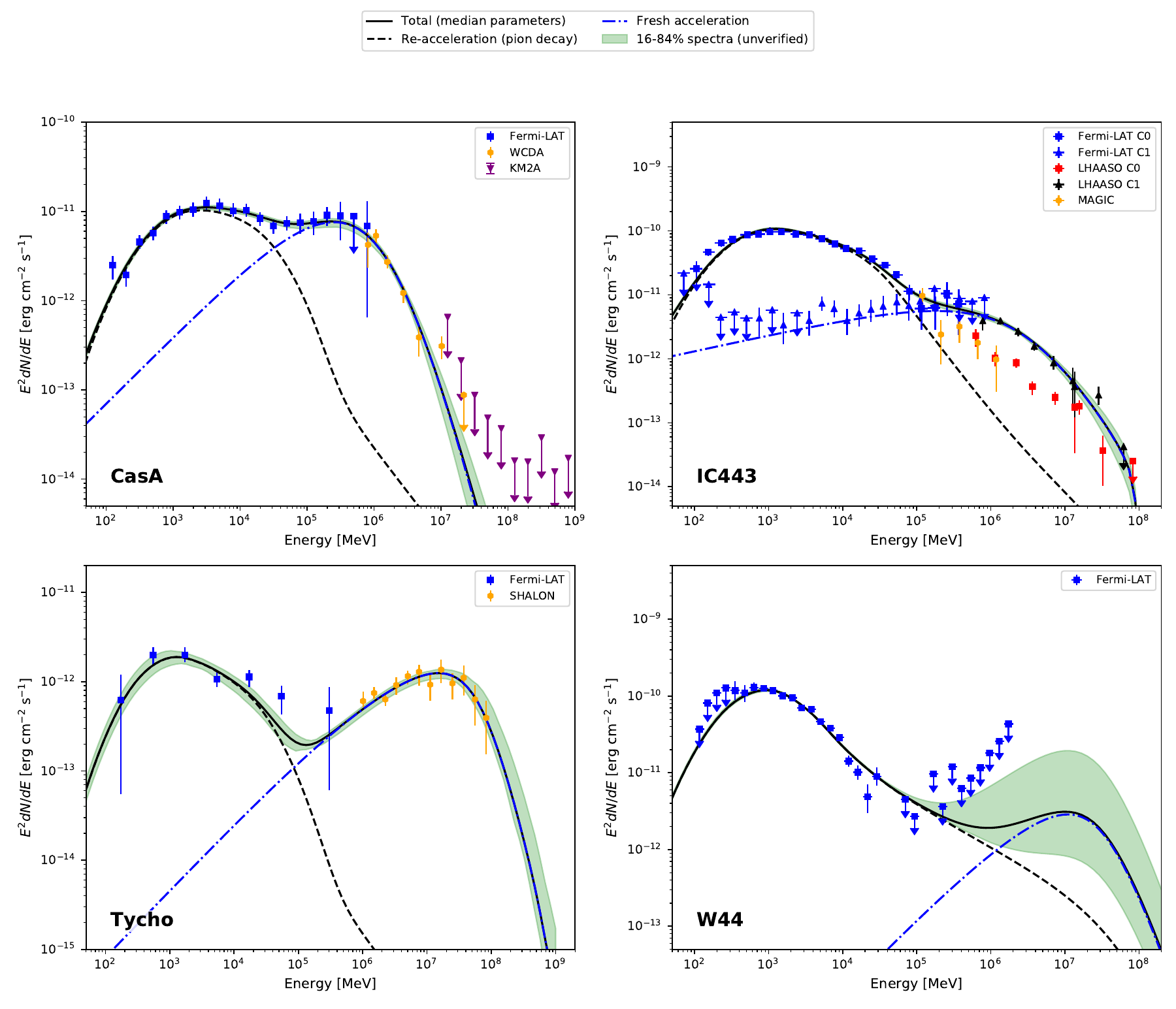}
\caption{Gamma-ray spectra of Cas A, IC 443, Tycho, and W44. The solid black curves show the total spectra calculated using the posterior median of each parameter. The dashed black and blue dash-dotted curves represent $\pi^0$-decay emission from reaccelerated protons and inverse Compton emission from freshly accelerated electrons, respectively. The green shaded bands indicate the 16th–84th percentile ranges of the model spectra corresponding to the parameter samples shown. The data take from the reference \citep{2025ApJ...982L..33C,2022ApJS..260...53A,2018AdSpR..62.2845S,2013Sci...339..807A,2007ApJ...664L..87A,2026PhRvL.136p1002C}.}
\label{fig1}
\end{figure}
\begin{figure}[t]%[htbp]
\centering
\includegraphics[width=0.98\linewidth]{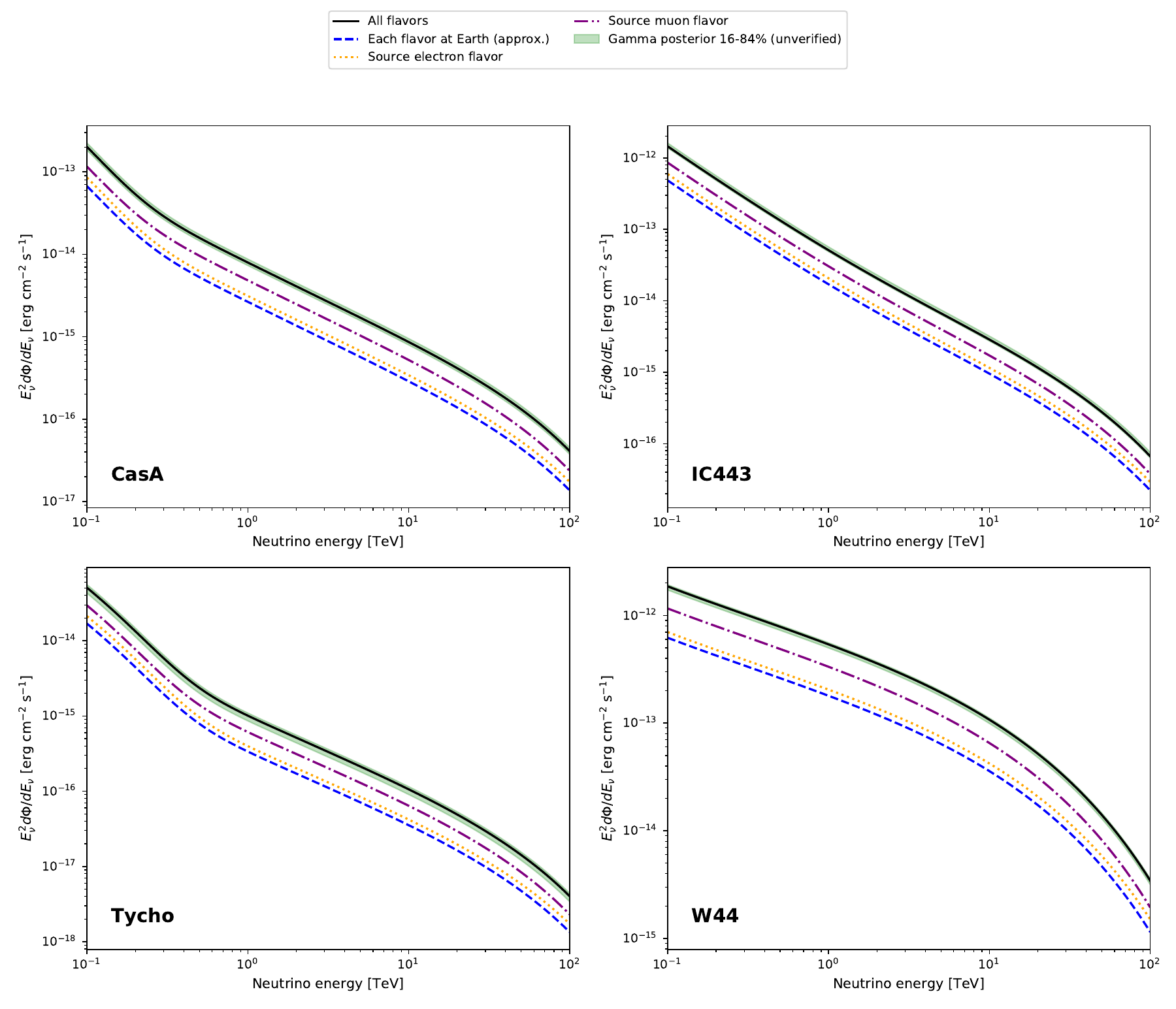}
\caption{Predicted neutrino spectra from $p$–$p$ interactions in Cas A, IC 443, Tycho, and W44. The solid black curves show the all-flavor fluxes. The orange dotted and purple dash-dotted curves represent the electron- and muon-flavor fluxes, respectively, including antineutrinos and geometric dilution but before flavor oscillations. The blue dash-dotted curves show the approximate per-flavor fluxes at Earth, estimated as one-third of the all-flavor fluxes. The green shaded bands indicate the 16th–84th percentile ranges of the all-flavor fluxes obtained by propagating the gamma-ray fit parameter samples.}
\label{fig2}
\end{figure}

\begin{figure}[t]%[htbp]
\centering
\includegraphics[width=0.98\linewidth]{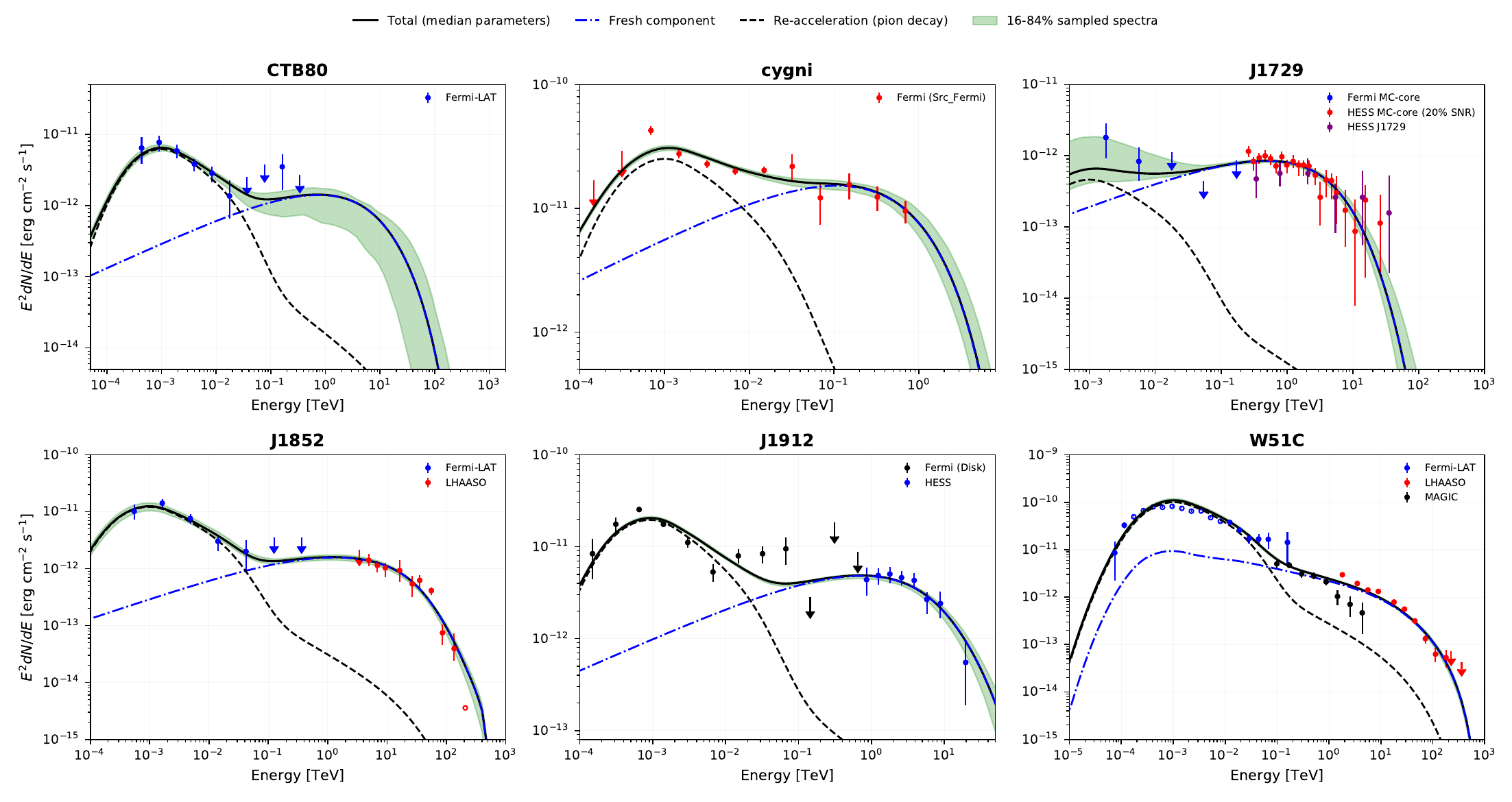}
\caption{The same as figure \ref{fig1} but for the source $\gamma$ Cygni, CTB 80, J1729, J1852, W51C, and J1912. The data take from the references \citep{2021MNRAS.502..472A,2025A&A...700A.143L,2011A&A...531A..81H,2019ApJ...887...47C,2026MNRAS.552g1719C,2018A&A...612A...8H,2023ApJ...953..100L,2016ApJ...816..100J,2012A&A...541A..13A,2024SciBu..69.2833C}.}
\label{fig3}
\end{figure}
\begin{figure}[h]%[htbp]
\centering
\includegraphics[width=0.98\linewidth]{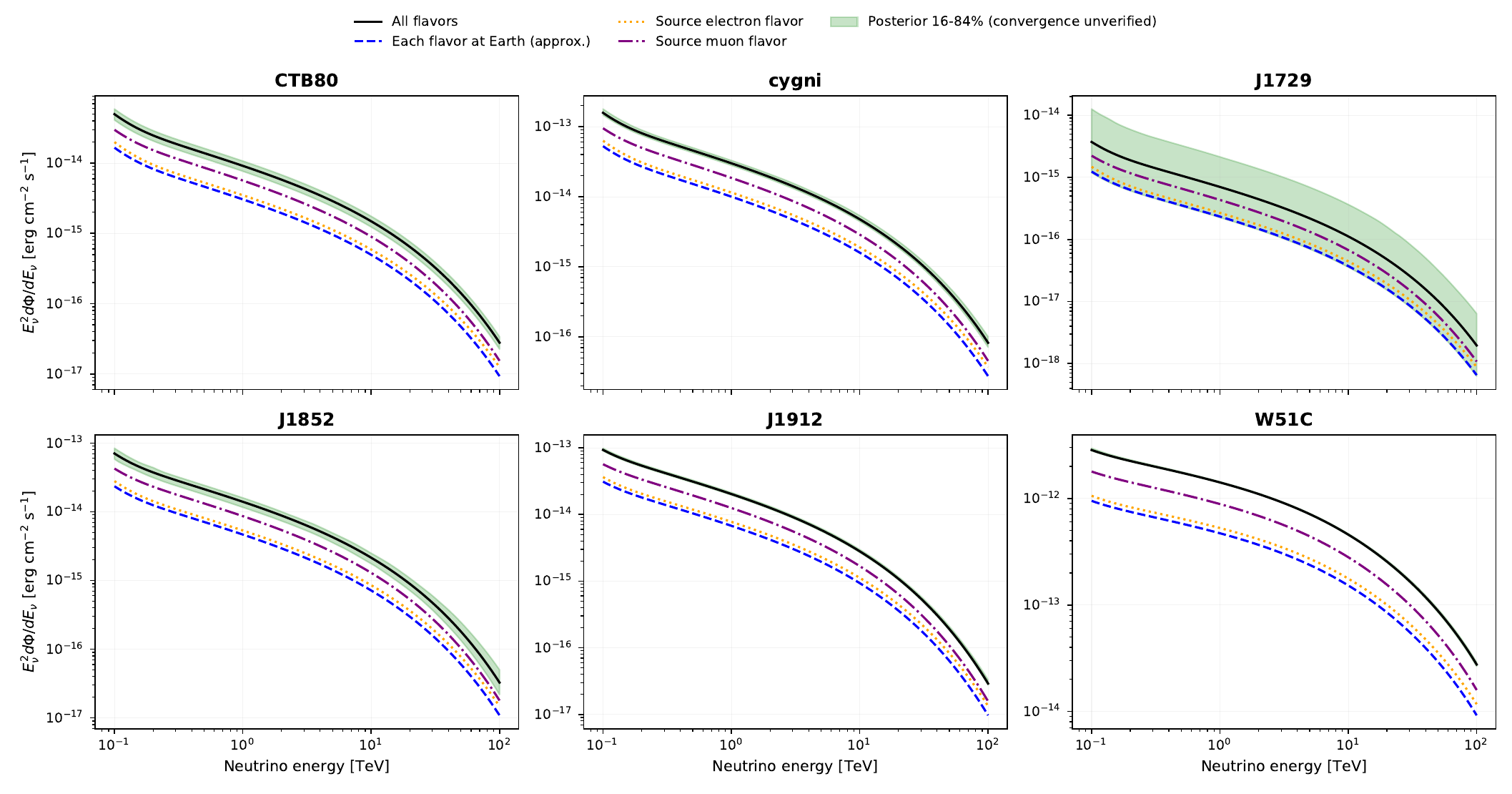}
\caption{The same as figure \ref{fig2} but for the source $\gamma$ Cygni, CTB 80, J1729, J1852, W51C, and J1912. }
\label{fig4}
\end{figure}

Paper I introduced the two-component model and applied it to Cas A, IC 443, Tycho, and W44, successfully reproducing their two-component spectral features and suggesting that such structures may be common among Galactic cosmic-ray sources. To facilitate interpretation and comparison, Figure \ref{fig1} reproduces the fits to these four sources, using the same free-parameter settings as in Paper I. Figure \ref{fig2} additionally presents the corresponding neutrino spectra from $p$–$p$ interactions. We emphasize that these spectra are theoretical predictions derived from proton models constrained by gamma-ray data, rather than fits to neutrino observations. The inverse Compton component does not produce the $p$–$p$ neutrinos calculated here. Consequently, the neutrino flux is not obtained by converting the observed total gamma-ray flux using a fixed proportionality factor.

With new observational data from LHAASO and other experiments, we extend the model to additional sources to enlarge the sample and assess the applicability of the two-component framework. The sources newly included in this work are $\gamma$ Cygni, CTB 80, J1729, J1852, W51C, and J1912. Figures \ref{fig3} and \ref{fig4} present the gamma-ray spectral fits and the corresponding $p$–$p$ neutrino predictions, respectively. Most sources exhibit a low-energy peak accompanied by a harder high-energy component, and some display two clearly separated spectral peaks. In the lepton-dominated scenario of our model, these components are described by $\pi^0$-decay emission from reaccelerated background protons and inverse Compton emission from freshly accelerated electrons, respectively. For sources with higher ambient gas densities, the high-energy component may instead be dominated by hadronic emission from freshly accelerated protons.

CTB 80 is a middle-aged supernova remnant with an estimated distance of $1.5$–$4.6\,\mathrm{kpc}$ \citep{2018ApJS..238...35S,2012MNRAS.423..718L}. We adopt $d=2\,\mathrm{kpc}$, a reference value commonly used in previous studies. At radio wavelengths, CTB 80 exhibits three bright, extended filamentary structures. CO and H I observations support the presence of a large, shock-driven shell associated with the remnant \citep{1990ApJ...364..178K,1993ApJ...417..196K}. LHAASO has detected a significant TeV emission excess near the eastern part of this radio shell \citep{2024ApJS..271...25C}, indicating very-high-energy particle activity in the region. Meanwhile, Fermi-LAT analyses have revealed significant extended GeV emission around CTB 80, with the brightest emission located along the northern radio arm. Previous observations also suggest that the remnant may be interacting with the surrounding medium in this region.

As shown in Figure \ref{fig3}, emission from reaccelerated background cosmic rays provides a good fit to the Fermi-LAT data for CTB 80. In our model, the low-energy component arises from interactions between dense gas and background cosmic rays that have undergone compression and reacceleration, whereas the emission at higher energies is supplied primarily by particles freshly accelerated by the source. As discussed in Paper I, both protons and electrons can be accelerated in low-density regions. However, if diffusion is suppressed within the remnant and acceleration occurs primarily near the forward shock, freshly accelerated protons generally need to escape the acceleration region and propagate to the relevant clouds before illuminating external dense gas. The contribution of these escaping particles must be treated consistently with the background cosmic-ray component to avoid double counting.

In dense regions, the shock decelerates substantially and may form a radiatively cooled, compressed layer, delaying the release of GeV particles. Given the small volume fraction occupied by dense gas, our baseline model neglects emission from freshly accelerated protons in these regions and assumes that their emission is dominated by reaccelerated background cosmic rays. In low-density regions, we primarily consider inverse Compton emission from freshly accelerated electrons. Thus, although some observations between GeV and sub-TeV energies currently provide only upper limits, the model predicts a possible additional inverse Compton component at higher energies in CTB 80. As more data accumulate, sensitive experiments such as LHAASO may further test this prediction.

W44 is a middle-aged, mixed-morphology supernova remnant with an age of approximately $20\,\mathrm{kyr}$ and a distance of approximately $2.2\,\mathrm{kpc}$. It exhibits a well-defined radio shell and bright thermal X-ray emission in its interior. The detection of OH (1720 MHz) masers provides direct evidence for interactions between the remnant shell and molecular clouds \citep{1997ApJ...489..143C}. Fermi-LAT has detected a characteristic $\pi^0$-decay “pion bump” in its low-energy gamma-ray spectrum \citep{2025A&A...693A.255A}, strongly supporting a hadronic origin of the emission. This feature alone, however, does not uniquely establish whether the emitting protons were freshly accelerated by the remnant or originated from reaccelerated background cosmic rays. Our model offers an alternative interpretation of the GeV emission from W44: background cosmic rays already present in nearby dense gas undergo compression and reacceleration as the expanding remnant shocks the molecular clouds, and subsequently produce gamma rays through $p$–$p$ interactions. In the low-density region occupying most of the remnant volume, the model assumes that emission is dominated by inverse Compton scattering from freshly accelerated electrons. As in CTB 80, we therefore predict a possible additional component at higher energies that awaits observational confirmation.

Several other sources discussed in this work already exhibit pronounced two-component spectral features. HESS J1731-347 is a young supernova remnant with nonthermal X-ray and TeV shell structures. A molecular cloud at a distance of approximately $3.2\,\mathrm{kpc}$ overlaps the western part of the remnant in projection and may be undergoing interaction with the remnant shock \citep{2017A&A...608A..23D,2018MNRAS.474..662M}. Localized GeV emission with a soft spectrum has also been detected coincident with the molecular cloud. We divide the surrounding environment into the MC-core region, the MC-J1729 region, and the low-density region outside these two dense molecular clumps. The latter primarily corresponds to the low-density environment shaped by the progenitor's stellar wind and its surrounding structures.

Figure \ref{fig3} presents the spectral fits for this region, with the MC-core and HESS J1729 labels identifying observational data from different regions. In our model, the low density of the wind bubble hosting the remnant implies that most of the region remains tenuous, apart from a small amount of dense gas. Its high-energy emission is therefore dominated by leptonic radiation from electrons accelerated by the remnant. The softer GeV component detected by Fermi-LAT arises primarily from the molecular clumps, where background cosmic rays are compressed and reaccelerated by the remnant shock and subsequently produce gamma rays through $p$–$p$ interactions.

This interpretation is compatible with suppressed diffusion around the source. Previous studies have suggested that describing particle transport near the remnant may require a diffusion coefficient approximately $3$–$20$ times smaller than the conventional Galactic value \citep{2007ApJ...665L.131G,2008A&A...481..401A}, and this picture has received support from various observations \citep{1981MNRAS.194..809L,2014A&A...567A..10L}. A lower diffusion coefficient can prolong particle residence times within molecular clumps. With a sustained particle supply and appropriate boundary conditions, this may enhance the local accumulation of cosmic rays, although the magnitude of the enhancement is determined by the transport process.

LHAASO has detected emission above $100\,\mathrm{TeV}$ from the extended source 1LHAASO J1852+0050u, which was identified as a new TeV source because of its substantial spatial extent \citep{2026MNRAS.552g1719C}. Its nearest pulsar is PSR J1853+0056, at a distance of $3.84\,\mathrm{kpc}$ and with a characteristic age of $204\,\mathrm{kyr}$, suggesting that the source may be a TeV halo associated with this pulsar. Alternatively, the extended GeV source Kes 79, identified by Fermi-LAT as potentially associated with a supernova remnant or a pulsar wind nebula, is also consistent with 1LHAASO J1852+0050u in position and extent. The extended source overlaps multiple H II regions in projection. CO-line observations indicate the presence of a molecular clump potentially associated with the gamma-ray source and linked to the ultracompact H II region G34.26+0.15 at a distance of approximately $3.3\,\mathrm{kpc}$ \citep{1994ApJ...436..117K}. The molecular-cloud gas density in this region is approximately $360\,\mathrm{cm^{-3}}$. Figure \ref{fig3} shows the two-component model fit to the observations.% and the corresponding model parameters are listed in Table 1.

HESS J1912+101 is a TeV source first discovered by the H.E.S.S. Collaboration in 2008 \citep{2008A&A...484..435A}. Its shell-like emission morphology makes it a supernova remnant candidate. The gas distribution in this region exhibits a good spatial correspondence with the gamma-ray intensity. In particular, the GeV emission closely coincides with dense molecular-cloud regions, suggesting that dense gas may provide important target material for the GeV emission. As shown in Figure \ref{fig3}, our model provides a good fit to the observed spectrum. Motivated by the spatial association between the molecular clouds and GeV emission, we interpret the GeV component as arising from background cosmic rays within the clouds that undergo reacceleration and compression following shock–cloud interactions and subsequently produce gamma rays through $p$–$p$ interactions. This interpretation is consistent with the observed spatial correlation.

The supernova remnant $\gamma$ Cygni (G78.2+2.1) lies in the complex Cygnus region. Despite extensive multiwavelength studies, the origin of its gamma-ray emission remains incompletely understood. The remnant has an estimated age of $7\,\mathrm{kyr}$ and a distance of approximately $1.7\,\mathrm{kpc}$ \citep{2000AstL...26...77L}. Its radio shell has an angular radius of approximately $0.5^\circ$, corresponding to a physical radius of approximately $15\,\mathrm{pc}$ \citep{2012A&A...542A...3S}. At TeV energies, the emission becomes less extended with increasing energy and progressively concentrates toward the remnant center. Since molecular clumps overlap portions of the GeV emission and radio shell in projection, we attribute the softer GeV component to compressed and reaccelerated background cosmic rays in the molecular clouds, while assigning the harder component at higher energies to electrons accelerated by the remnant itself. As shown in Figure \ref{fig3}, the combination of inverse Compton emission and the low-energy hadronic component provides a good description of the observed spectrum.

W51C is a supernova remnant identified through its radio shell, at a distance of approximately $5.5\,\mathrm{kpc}$ \citep{2010ApJ...720.1055S,2022MNRAS.516.6055R}. It belongs to the larger W51 complex, one of the largest star-forming regions in the Galaxy, which comprises the W51A and W51B regions. W51C lies at the southeastern end of W51B, and the detection of two OH (1720 MHz) masers confirms interactions between the remnant shell and the molecular cloud W51B \citep{1997AJ....114.2058G}. W51C is also among the few supernova remnants exhibiting a clear $\pi^0$-decay signature. Multiple observations indicate that its shock is interacting with dense clumps within the surrounding giant molecular cloud.
Both electrons and protons can be accelerated at the shock front of W51C, with electrons producing inverse Compton emission by scattering background photons. However, when the source environment contains both localized dense clumps and relatively high gas densities over a larger volume, hadronic emission from freshly accelerated protons may dominate the high-energy component. The density of the expanding atomic hydrogen gas driven by the remnant shock in W51C is estimated to be $30$–$160\,\mathrm{cm^{-3}}$ \citep{2024SciBu..69.2833C}. We therefore adopt a hadron-dominated interpretation of its high-energy emission in the baseline model and, following previous studies, assume an average gas density of $n_{\mathrm H}=100\,\mathrm{cm^{-3}}$.

As shown in Figures \ref{fig3} and \ref{fig4}, this model provides a good description of the Fermi-LAT and LHAASO data for W51C. The additional component in the Fermi-LAT energy range arises primarily from $p$–$p$ emission produced by reaccelerated and compressed background cosmic rays in dense clumps, whereas the LHAASO emission is dominated by hadronic interactions of protons freshly accelerated by the remnant. In this scenario, the high-energy gamma-ray component can appear as a relatively smooth shoulder, while the neutrino spectrum develops a harder component at the corresponding high energies owing to the freshly accelerated protons.

Figures \ref{fig2} and \ref{fig4} present the corresponding all-flavor neutrino fluxes. Our results show that sources evolving primarily in low-density environments but located near dense molecular clumps can exhibit pronounced two-component gamma-ray spectra. However, because the high-energy component is dominated by electron inverse Compton emission and does not contribute to the $p$–$p$ neutrinos calculated here, their neutrino emission primarily traces the reaccelerated background-proton component and is expected to have a relatively low flux. In contrast, for sources such as W51C, where relatively high gas densities extend over a larger region, gamma-ray emission above TeV energies can be dominated by hadronic processes. Both reaccelerated background protons and freshly accelerated protons then contribute to the neutrino emission, potentially producing a two-component neutrino spectrum with a comparatively higher flux at high energies. Such sources are therefore promising candidates for future searches for Galactic point-source neutrinos. Their actual detectability, however, requires further assessment accounting for the predicted flux, source extension, and detector sensitivity.

\section{Conclusion and Summary} \label{sec:conclusion}
Recent observations have revealed several phenomena that warrant further investigation: the spectra of different cosmic-ray “messengers” commonly exhibit two-component features, and similar structures have been observed in the gamma-ray spectra of an increasing number of sources located in dense environments. Whether these phenomena share a common physical origin is the central question motivating our study. In our previous work, we proposed a spatially dependent cosmic-ray propagation model to account for the two-component structure of cosmic-ray spectra, successfully explaining the spectral hardening of different cosmic-ray messengers at high energies. We subsequently extended this physical picture to individual sources and developed a two-component emission model, which provided a self-consistent explanation for the two-component spectral features revealed by LHAASO observations of sources such as Cas A and IC 443.

In this work, we apply the two-component model to additional sources with recently available observational data to further assess its applicability. The model predicts that, as more data accumulate, some sources, such as CTB 80 and J1857, may reveal spectral bumps at higher energies. In this framework, the low-energy bump originates primarily from background cosmic rays in dense molecular clouds. These particles undergo compression and reacceleration driven by the source and subsequently interact with the surrounding gas to produce gamma rays. The high-energy bump is instead produced primarily by freely propagating particles accelerated by the source, through inverse Compton scattering for electrons or proton–proton ($p$–$p$) interactions for protons.

Our results show that the ambient environment significantly affects the morphology of the two-component spectrum. For sources located in low-density environments but surrounded by dense molecular clumps, the two emission components produce clearly separated spectral peaks. For sources embedded in dense environments with additional dense molecular clumps nearby, the spectra retain two-component features, but the components overlap more smoothly. In these cases, the high-energy component appears primarily as a relatively broad shoulder, making two distinct peaks difficult to resolve. We also calculate the neutrino spectra of these sources, providing theoretical predictions for comparison with future neutrino observations.

The successful application of our two-component model to a larger sample of recently observed sources lends further support to the previously proposed picture of spatially dependent Galactic cosmic-ray propagation: high-energy cosmic rays are predominantly supplied by local sources, whereas low-energy cosmic rays arise mainly from the collective contribution of numerous sources in other regions of the Galaxy. Future observations of point-source and diffuse neutrino emission will offer further opportunities to test this physical picture from a multi-messenger perspective and place tighter constraints on the model parameters.

%\appendix
%\section{Some title}
%Please always give a title also for appendices.

\acknowledgments

This work is Supported by National Key R\&D Program of China [Grant No. 2024YFE0109900], the National Natural Science Foundation of China [Grant No. 12403097, 12333006, 12275279].

% Bibliography

%% [A] Recommended: using JHEP.bst file
 \bibliographystyle{JHEP}
 \bibliography{biblio.bib}

@PREAMBLE{
 "\providecommand{\noopsort}[1]{}" 
 # "\providecommand{\singleletter}[1]{#1}%" 
}

@ARTICLE{2015PhRvL.114q1103A,
       author = {{Aguilar}, M. and {Aisa}, D. and {Alpat}, B. and {Alvino}, A. and {Ambrosi}, G. and {Andeen}, K. and {Arruda}, L. and {Attig}, N. and {Azzarello}, P. and {Bachlechner}, A. and {Barao}, F. and {Barrau}, A. and {Barrin}, L. and {Bartoloni}, A. and {Basara}, L. and {Battarbee}, M. and {Battiston}, R. and {Bazo}, J. and {Becker}, U. and {Behlmann}, M. and {Beischer}, B. and {Berdugo}, J. and {Bertucci}, B. and {Bigongiari}, G. and {Bindi}, V. and {Bizzaglia}, S. and {Bizzarri}, M. and {Boella}, G. and {de Boer}, W. and {Bollweg}, K. and {Bonnivard}, V. and {Borgia}, B. and {Borsini}, S. and {Boschini}, M.~J. and {Bourquin}, M. and {Burger}, J. and {Cadoux}, F. and {Cai}, X.~D. and {Capell}, M. and {Caroff}, S. and {Casaus}, J. and {Cascioli}, V. and {Castellini}, G. and {Cernuda}, I. and {Cerreta}, D. and {Cervelli}, F. and {Chae}, M.~J. and {Chang}, Y.~H. and {Chen}, A.~I. and {Chen}, H. and {Cheng}, G.~M. and {Chen}, H.~S. and {Cheng}, L. and {Chou}, H.~Y. and {Choumilov}, E. and {Choutko}, V. and {Chung}, C.~H. and {Clark}, C. and {Clavero}, R. and {Coignet}, G. and {Consolandi}, C. and {Contin}, A. and {Corti}, C. and {Gil}, E. Cortina and {Coste}, B. and {Creus}, W. and {Crispoltoni}, M. and {Cui}, Z. and {Dai}, Y.~M. and {Delgado}, C. and {Della Torre}, S. and {Demirk{\"o}z}, M.~B. and {Derome}, L. and {Di Falco}, S. and {Di Masso}, L. and {Dimiccoli}, F. and {D{\'\i}az}, C. and {von Doetinchem}, P. and {Donnini}, F. and {Du}, W.~J. and {Duranti}, M. and {D'Urso}, D. and {Eline}, A. and {Eppling}, F.~J. and {Eronen}, T. and {Fan}, Y.~Y. and {Farnesini}, L. and {Feng}, J. and {Fiandrini}, E. and {Fiasson}, A. and {Finch}, E. and {Fisher}, P. and {Galaktionov}, Y. and {Gallucci}, G. and {Garc{\'\i}a}, B. and {Garc{\'\i}a-L{\'o}pez}, R. and {Gargiulo}, C. and {Gast}, H. and {Gebauer}, I. and {Gervasi}, M. and {Ghelfi}, A. and {Gillard}, W. and {Giovacchini}, F. and {Goglov}, P. and {Gong}, J. and {Goy}, C. and {Grabski}, V. and {Grandi}, D. and {Graziani}, M. and {Guandalini}, C. and {Guerri}, I. and {Guo}, K.~H. and {Haas}, D. and {Habiby}, M. and {Haino}, S. and {Han}, K.~C. and {He}, Z.~H. and {Heil}, M. and {Hoffman}, J. and {Hsieh}, T.~H. and {Huang}, Z.~C. and {Huh}, C. and {Incagli}, M. and {Ionica}, M. and {Jang}, W.~Y. and {Jinchi}, H. and {Kanishev}, K. and {Kim}, G.~N. and {Kim}, K.~S. and {Kirn}, Th. and {Kossakowski}, R. and {Kounina}, O. and {Kounine}, A. and {Koutsenko}, V. and {Krafczyk}, M.~S. and {La Vacca}, G. and {Laudi}, E. and {Laurenti}, G. and {Lazzizzera}, I. and {Lebedev}, A. and {Lee}, H.~T. and {Lee}, S.~C. and {Leluc}, C. and {Levi}, G. and {Li}, H.~L. and {Li}, J.~Q. and {Li}, Q. and {Li}, Q. and {Li}, T.~X. and {Li}, W. and {Li}, Y. and {Li}, Z.~H. and {Li}, Z.~Y. and {Lim}, S. and {Lin}, C.~H. and {Lipari}, P. and {Lippert}, T. and {Liu}, D. and {Liu}, H. and {Lolli}, M. and {Lomtadze}, T. and {Lu}, M.~J. and {Lu}, S.~Q. and {Lu}, Y.~S. and {Luebelsmeyer}, K. and {Luo}, J.~Z. and {Lv}, S.~S. and {Majka}, R. and {Ma{\~n}{\'a}}, C. and {Mar{\'\i}n}, J. and {Martin}, T. and {Mart{\'\i}nez}, G. and {Masi}, N. and {Maurin}, D. and {Menchaca-Rocha}, A. and {Meng}, Q. and {Mo}, D.~C. and {Morescalchi}, L. and {Mott}, P. and {M{\"u}ller}, M. and {Ni}, J.~Q. and {Nikonov}, N. and {Nozzoli}, F. and {Nunes}, P. and {Obermeier}, A. and {Oliva}, A. and {Orcinha}, M. and {Palmonari}, F. and {Palomares}, C. and {Paniccia}, M. and {Papi}, A. and {Pauluzzi}, M. and {Pedreschi}, E. and {Pensotti}, S. and {Pereira}, R. and {Picot-Clemente}, N. and {Pilo}, F. and {Piluso}, A. and {Pizzolotto}, C. and {Plyaskin}, V.},
        title = "{Precision Measurement of the Proton Flux in Primary Cosmic Rays from Rigidity 1 GV to 1.8 TV with the Alpha Magnetic Spectrometer on the International Space Station}",
      journal = {prl},
         year = 2015,
        month = may,
       volume = {114},
       number = {17},
          eid = {171103},
        pages = {171103},
          doi = {10.1103/PhysRevLett.114.171103},
       adsurl = {https://ui.adsabs.harvard.edu/abs/2015PhRvL.114q1103A}
}

@ARTICLE{2015PhRvL.115u1101A,
       author = {{Aguilar}, M. and {Aisa}, D. and {Alpat}, B. and {Alvino}, A. and {Ambrosi}, G. and {Andeen}, K. and {Arruda}, L. and {Attig}, N. and {Azzarello}, P. and {Bachlechner}, A. and {Barao}, F. and {Barrau}, A. and {Barrin}, L. and {Bartoloni}, A. and {Basara}, L. and {Battarbee}, M. and {Battiston}, R. and {Bazo}, J. and {Becker}, U. and {Behlmann}, M. and {Beischer}, B. and {Berdugo}, J. and {Bertucci}, B. and {Bindi}, V. and {Bizzaglia}, S. and {Bizzarri}, M. and {Boella}, G. and {de Boer}, W. and {Bollweg}, K. and {Bonnivard}, V. and {Borgia}, B. and {Borsini}, S. and {Boschini}, M.~J. and {Bourquin}, M. and {Burger}, J. and {Cadoux}, F. and {Cai}, X.~D. and {Capell}, M. and {Caroff}, S. and {Casaus}, J. and {Castellini}, G. and {Cernuda}, I. and {Cerreta}, D. and {Cervelli}, F. and {Chae}, M.~J. and {Chang}, Y.~H. and {Chen}, A.~I. and {Chen}, G.~M. and {Chen}, H. and {Chen}, H.~S. and {Cheng}, L. and {Chou}, H.~Y. and {Choumilov}, E. and {Choutko}, V. and {Chung}, C.~H. and {Clark}, C. and {Clavero}, R. and {Coignet}, G. and {Consolandi}, C. and {Contin}, A. and {Corti}, C. and {Gil}, E. Cortina and {Coste}, B. and {Creus}, W. and {Crispoltoni}, M. and {Cui}, Z. and {Dai}, Y.~M. and {Delgado}, C. and {Della Torre}, S. and {Demirk{\"o}z}, M.~B. and {Derome}, L. and {Di Falco}, S. and {Di Masso}, L. and {Dimiccoli}, F. and {D{\'\i}az}, C. and {von Doetinchem}, P. and {Donnini}, F. and {Duranti}, M. and {D'Urso}, D. and {Egorov}, A. and {Eline}, A. and {Eppling}, F.~J. and {Eronen}, T. and {Fan}, Y.~Y. and {Farnesini}, L. and {Feng}, J. and {Fiandrini}, E. and {Fiasson}, A. and {Finch}, E. and {Fisher}, P. and {Formato}, V. and {Galaktionov}, Y. and {Gallucci}, G. and {Garc{\'\i}a}, B. and {Garc{\'\i}a-L{\'o}pez}, R. and {Gargiulo}, C. and {Gast}, H. and {Gebauer}, I. and {Gervasi}, M. and {Ghelfi}, A. and {Giovacchini}, F. and {Goglov}, P. and {Gong}, J. and {Goy}, C. and {Grabski}, V. and {Grandi}, D. and {Graziani}, M. and {Guandalini}, C. and {Guerri}, I. and {Guo}, K.~H. and {Haas}, D. and {Habiby}, M. and {Haino}, S. and {Han}, K.~C. and {He}, Z.~H. and {Heil}, M. and {Hoffman}, J. and {Hsieh}, T.~H. and {Huang}, Z.~C. and {Huh}, C. and {Incagli}, M. and {Ionica}, M. and {Jang}, W.~Y. and {Jinchi}, H. and {Kanishev}, K. and {Kim}, G.~N. and {Kim}, K.~S. and {Kirn}, Th. and {Korkmaz}, M.~A. and {Kossakowski}, R. and {Kounina}, O. and {Kounine}, A. and {Koutsenko}, V. and {Krafczyk}, M.~S. and {La Vacca}, G. and {Laudi}, E. and {Laurenti}, G. and {Lazzizzera}, I. and {Lebedev}, A. and {Lee}, H.~T. and {Lee}, S.~C. and {Leluc}, C. and {Li}, H.~L. and {Li}, J.~Q. and {Li}, J.~Q. and {Li}, Q. and {Li}, Q. and {Li}, T.~X. and {Li}, W. and {Li}, Y. and {Li}, Z.~H. and {Li}, Z.~Y. and {Lim}, S. and {Lin}, C.~H. and {Lipari}, P. and {Lippert}, T. and {Liu}, D. and {Liu}, H. and {Liu}, Hu and {Lolli}, M. and {Lomtadze}, T. and {Lu}, M.~J. and {Lu}, S.~Q. and {Lu}, Y.~S. and {Luebelsmeyer}, K. and {Luo}, F. and {Luo}, J.~Z. and {Lv}, S.~S. and {Majka}, R. and {Ma{\~n}{\'a}}, C. and {Mar{\'\i}n}, J. and {Martin}, T. and {Mart{\'\i}nez}, G. and {Masi}, N. and {Maurin}, D. and {Menchaca-Rocha}, A. and {Meng}, Q. and {Mo}, D.~C. and {Morescalchi}, L. and {Mott}, P. and {M{\"u}ller}, M. and {Nelson}, T. and {Ni}, J.~Q. and {Nikonov}, N. and {Nozzoli}, F. and {Nunes}, P. and {Obermeier}, A. and {Oliva}, A. and {Orcinha}, M. and {Palmonari}, F. and {Palomares}, C. and {Paniccia}, M. and {Papi}, A. and {Pauluzzi}, M. and {Pedreschi}, E. and {Pensotti}, S. and {Pereira}, R. and {Picot-Clemente}, N. and {Pilo}, F. and {Piluso}, A.},
        title = "{Precision Measurement of the Helium Flux in Primary Cosmic Rays of Rigidities 1.9 GV to 3 TV with the Alpha Magnetic Spectrometer on the International Space Station}",
      journal = {prl},
         year = 2015,
        month = nov,
       volume = {115},
       number = {21},
          eid = {211101},
        pages = {211101},
          doi = {10.1103/PhysRevLett.115.211101},
       adsurl = {https://ui.adsabs.harvard.edu/abs/2015PhRvL.115u1101A}
}

@ARTICLE{2017PhRvL.119y1101A,
       author = {{Aguilar}, M. and {Ali Cavasonza}, L. and {Alpat}, B. and {Ambrosi}, G. and {Arruda}, L. and {Attig}, N. and {Aupetit}, S. and {Azzarello}, P. and {Bachlechner}, A. and {Barao}, F. and {Barrau}, A. and {Barrin}, L. and {Bartoloni}, A. and {Basara}, L. and {Ba{\textcommabelow s}e{\v{g}}mez-du Pree}, S. and {Battarbee}, M. and {Battiston}, R. and {Becker}, U. and {Behlmann}, M. and {Beischer}, B. and {Berdugo}, J. and {Bertucci}, B. and {Bindel}, K.~F. and {Bindi}, V. and {de Boer}, W. and {Bollweg}, K. and {Bonnivard}, V. and {Borgia}, B. and {Boschini}, M.~J. and {Bourquin}, M. and {Bueno}, E.~F. and {Burger}, J. and {Burger}, W.~J. and {Cadoux}, F. and {Cai}, X.~D. and {Capell}, M. and {Caroff}, S. and {Casaus}, J. and {Castellini}, G. and {Cervelli}, F. and {Chae}, M.~J. and {Chang}, Y.~H. and {Chen}, A.~I. and {Chen}, G.~M. and {Chen}, H.~S. and {Cheng}, L. and {Chou}, H.~Y. and {Choumilov}, E. and {Choutko}, V. and {Chung}, C.~H. and {Clark}, C. and {Clavero}, R. and {Coignet}, G. and {Consolandi}, C. and {Contin}, A. and {Corti}, C. and {Creus}, W. and {Crispoltoni}, M. and {Cui}, Z. and {Dadzie}, K. and {Dai}, Y.~M. and {Datta}, A. and {Delgado}, C. and {Della Torre}, S. and {Demakov}, O. and {Demirk{\"o}z}, M.~B. and {Derome}, L. and {Di Falco}, S. and {Dimiccoli}, F. and {D{\'\i}az}, C. and {von Doetinchem}, P. and {Dong}, F. and {Donnini}, F. and {Duranti}, M. and {D'Urso}, D. and {Egorov}, A. and {Eline}, A. and {Eronen}, T. and {Feng}, J. and {Fiandrini}, E. and {Fisher}, P. and {Formato}, V. and {Galaktionov}, Y. and {Gallucci}, G. and {Garc{\'\i}a-L{\'o}pez}, R.~J. and {Gargiulo}, C. and {Gast}, H. and {Gebauer}, I. and {Gervasi}, M. and {Ghelfi}, A. and {Giovacchini}, F. and {G{\'o}mez-Coral}, D.~M. and {Gong}, J. and {Goy}, C. and {Grabski}, V. and {Grandi}, D. and {Graziani}, M. and {Guo}, K.~H. and {Haino}, S. and {Han}, K.~C. and {He}, Z.~H. and {Heil}, M. and {Hoffman}, J. and {Hsieh}, T.~H. and {Huang}, H. and {Huang}, Z.~C. and {Huh}, C. and {Incagli}, M. and {Ionica}, M. and {Jang}, W.~Y. and {Jia}, Yi and {Jinchi}, H. and {Kang}, S.~C. and {Kanishev}, K. and {Khiali}, B. and {Kim}, G.~N. and {Kim}, K.~S. and {Kirn}, Th. and {Konak}, C. and {Kounina}, O. and {Kounine}, A. and {Koutsenko}, V. and {Kulemzin}, A. and {La Vacca}, G. and {Laudi}, E. and {Laurenti}, G. and {Lazzizzera}, I. and {Lebedev}, A. and {Lee}, H.~T. and {Lee}, S.~C. and {Leluc}, C. and {Li}, H.~S. and {Li}, J.~Q. and {Li}, Q. and {Li}, T.~X. and {Li}, Y. and {Li}, Z.~H. and {Li}, Z.~Y. and {Lim}, S. and {Lin}, C.~H. and {Lipari}, P. and {Lippert}, T. and {Liu}, D. and {Liu}, Hu and {Lordello}, V.~D. and {Lu}, S.~Q. and {Lu}, Y.~S. and {Luebelsmeyer}, K. and {Luo}, F. and {Luo}, J.~Z. and {Lyu}, S.~S. and {Machate}, F. and {Ma{\~n}{\'a}}, C. and {Mar{\'\i}n}, J. and {Martin}, T. and {Mart{\'\i}nez}, G. and {Masi}, N. and {Maurin}, D. and {Menchaca-Rocha}, A. and {Meng}, Q. and {Mikuni}, V.~M. and {Mo}, D.~C. and {Mott}, P. and {Nelson}, T. and {Ni}, J.~Q. and {Nikonov}, N. and {Nozzoli}, F. and {Oliva}, A. and {Orcinha}, M. and {Palmonari}, F. and {Palomares}, C. and {Paniccia}, M. and {Pauluzzi}, M. and {Pensotti}, S. and {Perrina}, C. and {Phan}, H.~D. and {Picot-Clemente}, N. and {Pilo}, F. and {Pizzolotto}, C. and {Plyaskin}, V. and {Pohl}, M. and {Poireau}, V. and {Quadrani}, L. and {Qi}, X.~M. and {Qin}, X. and {Qu}, Z.~Y. and {R{\"a}ih{\"a}}, T. and {Rancoita}, P.~G. and {Rapin}, D. and {Ricol}, J.~S. and {Rosier-Lees}, S. and {Rozhkov}, A. and {Rozza}, D. and {Sagdeev}, R. and {Schael}, S. and {Schmidt}, S.~M. and {Schulz von Dratzig}, A. and {Schwering}, G. and {Seo}, E.~S. and {Shan}, B.~S.},
        title = "{Observation of the Identical Rigidity Dependence of He, C, and O Cosmic Rays at High Rigidities by the Alpha Magnetic Spectrometer on the International Space Station}",
      journal = {prl},
         year = 2017,
        month = dec,
       volume = {119},
       number = {25},
          eid = {251101},
        pages = {251101},
          doi = {10.1103/PhysRevLett.119.251101},
       adsurl = {https://ui.adsabs.harvard.edu/abs/2017PhRvL.119y1101A}
}

@ARTICLE{2018PhRvL.120b1101A,
       author = {{Aguilar}, M. and {Ali Cavasonza}, L. and {Ambrosi}, G. and {Arruda}, L. and {Attig}, N. and {Aupetit}, S. and {Azzarello}, P. and {Bachlechner}, A. and {Barao}, F. and {Barrau}, A. and {Barrin}, L. and {Bartoloni}, A. and {Basara}, L. and {Ba{\textcommabelow s}e{\v{g}}mez-du Pree}, S. and {Battarbee}, M. and {Battiston}, R. and {Becker}, U. and {Behlmann}, M. and {Beischer}, B. and {Berdugo}, J. and {Bertucci}, B. and {Bindel}, K.~F. and {Bindi}, V. and {de Boer}, W. and {Bollweg}, K. and {Bonnivard}, V. and {Borgia}, B. and {Boschini}, M.~J. and {Bourquin}, M. and {Bueno}, E.~F. and {Burger}, J. and {Burger}, W.~J. and {Cadoux}, F. and {Cai}, X.~D. and {Capell}, M. and {Caroff}, S. and {Casaus}, J. and {Castellini}, G. and {Cervelli}, F. and {Chae}, M.~J. and {Chang}, Y.~H. and {Chen}, A.~I. and {Chen}, G.~M. and {Chen}, H.~S. and {Cheng}, L. and {Chou}, H.~Y. and {Choumilov}, E. and {Choutko}, V. and {Chung}, C.~H. and {Clark}, C. and {Clavero}, R. and {Coignet}, G. and {Consolandi}, C. and {Contin}, A. and {Corti}, C. and {Creus}, W. and {Crispoltoni}, M. and {Cui}, Z. and {Dadzie}, K. and {Dai}, Y.~M. and {Datta}, A. and {Delgado}, C. and {Della Torre}, S. and {Demirk{\"o}z}, M.~B. and {Derome}, L. and {Di Falco}, S. and {Dimiccoli}, F. and {D{\'\i}az}, C. and {von Doetinchem}, P. and {Dong}, F. and {Donnini}, F. and {Duranti}, M. and {D'Urso}, D. and {Egorov}, A. and {Eline}, A. and {Eronen}, T. and {Feng}, J. and {Fiandrini}, E. and {Fisher}, P. and {Formato}, V. and {Galaktionov}, Y. and {Gallucci}, G. and {Garc{\'\i}a-L{\'o}pez}, R.~J. and {Gargiulo}, C. and {Gast}, H. and {Gebauer}, I. and {Gervasi}, M. and {Ghelfi}, A. and {Giovacchini}, F. and {G{\'o}mez-Coral}, D.~M. and {Gong}, J. and {Goy}, C. and {Grabski}, V. and {Grandi}, D. and {Graziani}, M. and {Guo}, K.~H. and {Haino}, S. and {Han}, K.~C. and {He}, Z.~H. and {Heil}, M. and {Hsieh}, T.~H. and {Huang}, H. and {Huang}, Z.~C. and {Huh}, C. and {Incagli}, M. and {Ionica}, M. and {Jang}, W.~Y. and {Jia}, Yi and {Jinchi}, H. and {Kang}, S.~C. and {Kanishev}, K. and {Khiali}, B. and {Kim}, G.~N. and {Kim}, K.~S. and {Kirn}, Th. and {Konak}, C. and {Kounina}, O. and {Kounine}, A. and {Koutsenko}, V. and {Kulemzin}, A. and {La Vacca}, G. and {Laudi}, E. and {Laurenti}, G. and {Lazzizzera}, I. and {Lebedev}, A. and {Lee}, H.~T. and {Lee}, S.~C. and {Leluc}, C. and {Li}, H.~S. and {Li}, J.~Q. and {Li}, Q. and {Li}, T.~X. and {Li}, Y. and {Li}, Z.~H. and {Li}, Z.~Y. and {Lim}, S. and {Lin}, C.~H. and {Lipari}, P. and {Lippert}, T. and {Liu}, D. and {Liu}, Hu and {Lordello}, V.~D. and {Lu}, S.~Q. and {Lu}, Y.~S. and {Luebelsmeyer}, K. and {Luo}, F. and {Luo}, J.~Z. and {Lyu}, S.~S. and {Machate}, F. and {Ma{\~n}{\'a}}, C. and {Mar{\'\i}n}, J. and {Martin}, T. and {Mart{\'\i}nez}, G. and {Masi}, N. and {Maurin}, D. and {Menchaca-Rocha}, A. and {Meng}, Q. and {Mikuni}, V.~M. and {Mo}, D.~C. and {Mott}, P. and {Nelson}, T. and {Ni}, J.~Q. and {Nikonov}, N. and {Nozzoli}, F. and {Oliva}, A. and {Orcinha}, M. and {Palermo}, M. and {Palmonari}, F. and {Palomares}, C. and {Paniccia}, M. and {Pauluzzi}, M. and {Pensotti}, S. and {Perrina}, C. and {Phan}, H.~D. and {Picot-Clemente}, N. and {Pilo}, F. and {Pizzolotto}, C. and {Plyaskin}, V. and {Pohl}, M. and {Poireau}, V. and {Quadrani}, L. and {Qi}, X.~M. and {Qin}, X. and {Qu}, Z.~Y. and {R{\"a}ih{\"a}}, T. and {Rancoita}, P.~G. and {Rapin}, D. and {Ricol}, J.~S. and {Rosier-Lees}, S. and {Rozhkov}, A. and {Rozza}, D. and {Sagdeev}, R. and {Schael}, S. and {Schmidt}, S.~M. and {Schulz von Dratzig}, A. and {Schwering}, G. and {Seo}, E.~S. and {Shan}, B.~S. and {Shi}, J.~Y. and {Siedenburg}, T.},
        title = "{Observation of New Properties of Secondary Cosmic Rays Lithium, Beryllium, and Boron by the Alpha Magnetic Spectrometer on the International Space Station}",
      journal = {prl},
         year = 2018,
        month = jan,
       volume = {120},
       number = {2},
          eid = {021101},
        pages = {021101},
          doi = {10.1103/PhysRevLett.120.021101},
       adsurl = {https://ui.adsabs.harvard.edu/abs/2018PhRvL.120b1101A}
}

@ARTICLE{2018PhRvL.121e1103A,
       author = {{Aguilar}, M. and {Ali Cavasonza}, L. and {Alpat}, B. and {Ambrosi}, G. and {Arruda}, L. and {Attig}, N. and {Aupetit}, S. and {Azzarello}, P. and {Bachlechner}, A. and {Barao}, F. and {Barrau}, A. and {Barrin}, L. and {Bartoloni}, A. and {Basara}, L. and {Ba{\textcommabelow s}e{\v{g}}mez-du Pree}, S. and {Battarbee}, M. and {Battiston}, R. and {Becker}, U. and {Behlmann}, M. and {Beischer}, B. and {Berdugo}, J. and {Bertucci}, B. and {Bindel}, K.~F. and {Bindi}, V. and {de Boer}, W. and {Bollweg}, K. and {Bonnivard}, V. and {Borgia}, B. and {Boschini}, M.~J. and {Bourquin}, M. and {Bueno}, E.~F. and {Burger}, J. and {Burger}, W.~J. and {Cai}, X.~D. and {Capell}, M. and {Caroff}, S. and {Casaus}, J. and {Castellini}, G. and {Cervelli}, F. and {Chang}, Y.~H. and {Chen}, A.~I. and {Chen}, G.~M. and {Chen}, H.~S. and {Chen}, Y. and {Cheng}, L. and {Chou}, H.~Y. and {Choumilov}, E. and {Choutko}, V. and {Chung}, C.~H. and {Clark}, C. and {Clavero}, R. and {Coignet}, G. and {Consolandi}, C. and {Contin}, A. and {Corti}, C. and {Creus}, W. and {Crispoltoni}, M. and {Cui}, Z. and {Dadzie}, K. and {Dai}, Y.~M. and {Datta}, A. and {Delgado}, C. and {Della Torre}, S. and {Demirk{\"o}z}, M.~B. and {Derome}, L. and {Di Falco}, S. and {Dimiccoli}, F. and {D{\'\i}az}, C. and {von Doetinchem}, P. and {Dong}, F. and {Donnini}, F. and {Duranti}, M. and {Egorov}, A. and {Eline}, A. and {Eronen}, T. and {Feng}, J. and {Fiandrini}, E. and {Fisher}, P. and {Formato}, V. and {Galaktionov}, Y. and {Gallucci}, G. and {Garc{\'\i}a-L{\'o}pez}, R.~J. and {Gargiulo}, C. and {Gast}, H. and {Gebauer}, I. and {Gervasi}, M. and {Ghelfi}, A. and {Giovacchini}, F. and {G{\'o}mez-Coral}, D.~M. and {Gong}, J. and {Goy}, C. and {Grabski}, V. and {Grandi}, D. and {Graziani}, M. and {Guo}, K.~H. and {Haino}, S. and {Han}, K.~C. and {He}, Z.~H. and {Heil}, M. and {Hsieh}, T.~H. and {Huang}, H. and {Huang}, Z.~C. and {Incagli}, M. and {Jia}, Yi and {Jinchi}, H. and {Kanishev}, K. and {Khiali}, B. and {Kirn}, Th. and {Konak}, C. and {Kounina}, O. and {Kounine}, A. and {Koutsenko}, V. and {Kulemzin}, A. and {La Vacca}, G. and {Laudi}, E. and {Laurenti}, G. and {Lazzizzera}, I. and {Lebedev}, A. and {Lee}, H.~T. and {Lee}, S.~C. and {Leluc}, C. and {Li}, H.~S. and {Li}, J.~Q. and {Li}, Q. and {Li}, T.~X. and {Li}, Z.~H. and {Li}, Z.~Y. and {Lin}, C.~H. and {Lipari}, P. and {Lippert}, T. and {Liu}, D. and {Liu}, Hu and {Liu}, Z. and {Lordello}, V.~D. and {Lu}, S.~Q. and {Lu}, Y.~S. and {Luebelsmeyer}, K. and {Luo}, F. and {Luo}, J.~Z. and {Lyu}, S.~S. and {Machate}, F. and {Ma{\~n}{\'a}}, C. and {Mar{\'\i}n}, J. and {Martin}, T. and {Mart{\'\i}nez}, G. and {Masi}, N. and {Maurin}, D. and {Menchaca-Rocha}, A. and {Meng}, Q. and {Mikuni}, V.~M. and {Mo}, D.~C. and {Mott}, P. and {Mussolin}, L. and {Nelson}, T. and {Ni}, J.~Q. and {Nikonov}, N. and {Nozzoli}, F. and {Oliva}, A. and {Orcinha}, M. and {Palermo}, M. and {Palmonari}, F. and {Palomares}, C. and {Paniccia}, M. and {Pauluzzi}, M. and {Pensotti}, S. and {Perrina}, C. and {Phan}, H.~D. and {Picot-Clemente}, N. and {Pilo}, F. and {Plyaskin}, V. and {Pohl}, M. and {Poireau}, V. and {Quadrani}, L. and {Qi}, X.~M. and {Qin}, X. and {Qu}, Z.~Y. and {R{\"a}ih{\"a}}, T. and {Rancoita}, P.~G. and {Rapin}, D. and {Ricol}, J.~S. and {Rosier-Lees}, S. and {Rozhkov}, A. and {Rozza}, D. and {Sagdeev}, R. and {Schael}, S. and {Schmidt}, S.~M. and {Schulz von Dratzig}, A. and {Schwering}, G. and {Seo}, E.~S. and {Shan}, B.~S. and {Shi}, J.~Y. and {Siedenburg}, T. and {Song}, J.~W. and {Tacconi}, M. and {Tang}, X.~W. and {Tang}, Z.~C. and {Tescaro}, D. and {Tian}, J. and {Ting}, Samuel C.~C. and {Ting}, S.~M.},
        title = "{Precision Measurement of Cosmic-Ray Nitrogen and its Primary and Secondary Components with the Alpha Magnetic Spectrometer on the International Space Station}",
      journal = {prl},
         year = 2018,
        month = aug,
       volume = {121},
       number = {5},
          eid = {051103},
        pages = {051103},
          doi = {10.1103/PhysRevLett.121.051103},
       adsurl = {https://ui.adsabs.harvard.edu/abs/2018PhRvL.121e1103A}
}

@ARTICLE{2020PhRvL.124u1102A,
       author = {{Aguilar}, M. and {Ali Cavasonza}, L. and {Ambrosi}, G. and {Arruda}, L. and {Attig}, N. and {Barao}, F. and {Barrin}, L. and {Bartoloni}, A. and {Ba{\textcommabelow s}e{\v{g}}mez-du Pree}, S. and {Battiston}, R. and {Becker}, U. and {Behlmann}, M. and {Beischer}, B. and {Berdugo}, J. and {Bertucci}, B. and {Bindi}, V. and {de Boer}, W. and {Bollweg}, K. and {Borgia}, B. and {Boschini}, M.~J. and {Bourquin}, M. and {Bueno}, E.~F. and {Burger}, J. and {Burger}, W.~J. and {Burmeister}, S. and {Cai}, X.~D. and {Capell}, M. and {Casaus}, J. and {Castellini}, G. and {Cervelli}, F. and {Chang}, Y.~H. and {Chen}, G.~M. and {Chen}, H.~S. and {Chen}, Y. and {Cheng}, L. and {Chou}, H.~Y. and {Chouridou}, S. and {Choutko}, V. and {Chung}, C.~H. and {Clark}, C. and {Coignet}, G. and {Consolandi}, C. and {Contin}, A. and {Corti}, C. and {Cui}, Z. and {Dadzie}, K. and {Dai}, Y.~M. and {Delgado}, C. and {Della Torre}, S. and {Demirk{\"o}z}, M.~B. and {Derome}, L. and {Di Falco}, S. and {Di Felice}, V. and {D{\'\i}az}, C. and {Dimiccoli}, F. and {von Doetinchem}, P. and {Dong}, F. and {Donnini}, F. and {Duranti}, M. and {Egorov}, A. and {Eline}, A. and {Feng}, J. and {Fiandrini}, E. and {Fisher}, P. and {Formato}, V. and {Freeman}, C. and {Galaktionov}, Y. and {G{\'a}mez}, C. and {Garc{\'\i}a-L{\'o}pez}, R.~J. and {Gargiulo}, C. and {Gast}, H. and {Gebauer}, I. and {Gervasi}, M. and {Giovacchini}, F. and {G{\'o}mez-Coral}, D.~M. and {Gong}, J. and {Goy}, C. and {Grabski}, V. and {Grandi}, D. and {Graziani}, M. and {Guo}, K.~H. and {Haino}, S. and {Han}, K.~C. and {Hashmani}, R.~K. and {He}, Z.~H. and {Heber}, B. and {Hsieh}, T.~H. and {Hu}, J.~Y. and {Huang}, Z.~C. and {Incagli}, M. and {Jang}, W.~Y. and {Jia}, Yi and {Jinchi}, H. and {Kanishev}, K. and {Khiali}, B. and {Kim}, G.~N. and {Kirn}, Th. and {Konyushikhin}, M. and {Kounina}, O. and {Kounine}, A. and {Koutsenko}, V. and {Kuhlman}, A. and {Kulemzin}, A. and {La Vacca}, G. and {Laudi}, E. and {Laurenti}, G. and {Lazzizzera}, I. and {Lebedev}, A. and {Lee}, H.~T. and {Lee}, S.~C. and {Li}, J.~Q. and {Li}, M. and {Li}, Q. and {Li}, S. and {Li}, T.~X. and {Li}, Z.~H. and {Light}, C. and {Lin}, C.~H. and {Lippert}, T. and {Liu}, Z. and {Lu}, S.~Q. and {Lu}, Y.~S. and {Luebelsmeyer}, K. and {Luo}, J.~Z. and {Lyu}, S.~S. and {Machate}, F. and {Ma{\~n}{\'a}}, C. and {Mar{\'\i}n}, J. and {Marquardt}, J. and {Martin}, T. and {Mart{\'\i}nez}, G. and {Masi}, N. and {Maurin}, D. and {Menchaca-Rocha}, A. and {Meng}, Q. and {Mo}, D.~C. and {Molero}, M. and {Mott}, P. and {Mussolin}, L. and {Ni}, J.~Q. and {Nikonov}, N. and {Nozzoli}, F. and {Oliva}, A. and {Orcinha}, M. and {Palermo}, M. and {Palmonari}, F. and {Paniccia}, M. and {Pashnin}, A. and {Pauluzzi}, M. and {Pensotti}, S. and {Phan}, H.~D. and {Piandani}, R. and {Plyaskin}, V. and {Poluianov}, S. and {Qi}, X.~M. and {Qin}, X. and {Qu}, Z.~Y. and {Quadrani}, L. and {Rancoita}, P.~G. and {Rapin}, D. and {Reina Conde}, A. and {Rosier-Lees}, S. and {Rozhkov}, A. and {Rozza}, D. and {Sagdeev}, R. and {Schael}, S. and {Schmidt}, S.~M. and {Schulz von Dratzig}, A. and {Schwering}, G. and {Seo}, E.~S. and {Shan}, B.~S. and {Shi}, J.~Y. and {Siedenburg}, T. and {Solano}, C. and {Sonnabend}, R. and {Song}, J.~W. and {Sun}, Q. and {Sun}, Z.~T. and {Tacconi}, M. and {Tang}, X.~W. and {Tang}, Z.~C. and {Tian}, J. and {Ting}, Samuel C.~C. and {Ting}, S.~M. and {Tomassetti}, N. and {Torsti}, J. and {T{\"u}ys{\"u}z}, C. and {Urban}, T. and {Usoskin}, I. and {Vagelli}, V. and {Vainio}, R. and {Valente}, E. and {Valtonen}, E. and {V{\'a}zquez Acosta}, M. and {Vecchi}, M. and {Velasco}, M. and {Vialle}, J.~P. and {Wallmann}, C. and {Wang}, L.~Q. and {Wang}, N.~H.},
        title = "{Properties of Neon, Magnesium, and Silicon Primary Cosmic Rays Results from the Alpha Magnetic Spectrometer}",
      journal = {prl},
         year = 2020,
        month = may,
       volume = {124},
       number = {21},
          eid = {211102},
        pages = {211102},
          doi = {10.1103/PhysRevLett.124.211102},
       adsurl = {https://ui.adsabs.harvard.edu/abs/2020PhRvL.124u1102A}
}

@ARTICLE{2021PhRvL.126d1104A,
       author = {{Aguilar}, M. and {Cavasonza}, L. Ali and {Allen}, M.~S. and {Alpat}, B. and {Ambrosi}, G. and {Arruda}, L. and {Attig}, N. and {Barao}, F. and {Barrin}, L. and {Bartoloni}, A. and {Ba{\textcommabelow s}e{\v{g}}mez-du Pree}, S. and {Battiston}, R. and {Behlmann}, M. and {Beischer}, B. and {Berdugo}, J. and {Bertucci}, B. and {Bindi}, V. and {de Boer}, W. and {Bollweg}, K. and {Borgia}, B. and {Boschini}, M.~J. and {Bourquin}, M. and {Bueno}, E.~F. and {Burger}, J. and {Burger}, W.~J. and {Burmeister}, S. and {Cai}, X.~D. and {Capell}, M. and {Casaus}, J. and {Castellini}, G. and {Cervelli}, F. and {Chang}, Y.~H. and {Chen}, G.~M. and {Chen}, G.~R. and {Chen}, H.~S. and {Chen}, Y. and {Cheng}, L. and {Chou}, H.~Y. and {Chouridou}, S. and {Choutko}, V. and {Chung}, C.~H. and {Clark}, C. and {Coignet}, G. and {Consolandi}, C. and {Contin}, A. and {Corti}, C. and {Cui}, Z. and {Dadzie}, K. and {Delgado}, C. and {Della Torre}, S. and {Demirk{\"o}z}, M.~B. and {Derome}, L. and {Di Falco}, S. and {Di Felice}, V. and {D{\'\i}az}, C. and {Dimiccoli}, F. and {von Doetinchem}, P. and {Dong}, F. and {Donnini}, F. and {Duranti}, M. and {Egorov}, A. and {Eline}, A. and {Feng}, J. and {Fiandrini}, E. and {Fisher}, P. and {Formato}, V. and {Freeman}, C. and {Galaktionov}, Y. and {G{\'a}mez}, C. and {Garc{\'\i}a-L{\'o}pez}, R.~J. and {Gargiulo}, C. and {Gast}, H. and {Gervasi}, M. and {Giovacchini}, F. and {G{\'o}mez-Coral}, D.~M. and {Gong}, J. and {Goy}, C. and {Grabski}, V. and {Grandi}, D. and {Graziani}, M. and {Haino}, S. and {Han}, K.~C. and {Hashmani}, R.~K. and {He}, Z.~H. and {Heber}, B. and {Hsieh}, T.~H. and {Hu}, J.~Y. and {Incagli}, M. and {Jang}, W.~Y. and {Jia}, Yi and {Jinchi}, H. and {Kanishev}, K. and {Khiali}, B. and {Kim}, G.~N. and {Kirn}, Th. and {Konyushikhin}, M. and {Kounina}, O. and {Kounine}, A. and {Koutsenko}, V. and {Kuhlman}, A. and {Kulemzin}, A. and {La Vacca}, G. and {Laudi}, E. and {Laurenti}, G. and {Lazzizzera}, I. and {Lebedev}, A. and {Lee}, H.~T. and {Lee}, S.~C. and {Li}, J.~Q. and {Li}, M. and {Li}, Q. and {Li}, S. and {Li}, J.~H. and {Li}, Z.~H. and {Liang}, J. and {Light}, C. and {Lin}, C.~H. and {Lippert}, T. and {Liu}, J.~H. and {Liu}, Z. and {Lu}, S.~Q. and {Lu}, Y.~S. and {Luebelsmeyer}, K. and {Luo}, J.~Z. and {Luo}, Xi and {Lyu}, S.~S. and {Machate}, F. and {Ma{\~n}{\'a}}, C. and {Mar{\'\i}n}, J. and {Marquardt}, J. and {Martin}, T. and {Mart{\'\i}nez}, G. and {Masi}, N. and {Maurin}, D. and {Menchaca-Rocha}, A. and {Meng}, Q. and {Mikhailov}, V.~V. and {Mo}, D.~C. and {Molero}, M. and {Mott}, P. and {Mussolin}, L. and {Negrete}, J. and {Nikonov}, N. and {Nozzoli}, F. and {Oliva}, A. and {Orcinha}, M. and {Palermo}, M. and {Palmonari}, F. and {Paniccia}, M. and {Pashnin}, A. and {Pauluzzi}, M. and {Pensotti}, S. and {Phan}, H.~D. and {Piandani}, R. and {Plyaskin}, V. and {Poluianov}, S. and {Qin}, X. and {Qu}, Z.~Y. and {Quadrani}, L. and {Rancoita}, P.~G. and {Rapin}, D. and {Conde}, A. Reina and {Robyn}, E. and {Rosier-Lees}, S. and {Rozhkov}, A. and {Rozza}, D. and {Sagdeev}, R. and {Schael}, S. and {von Dratzig}, A. Schulz and {Schwering}, G. and {Seo}, E.~S. and {Shakfa}, Z. and {Shan}, B.~S. and {Siedenburg}, T. and {Solano}, C. and {Song}, J.~W. and {Song}, X.~J. and {Sonnabend}, R. and {Strigari}, L. and {Su}, T. and {Sun}, Q. and {Sun}, Z.~T. and {Tacconi}, M. and {Tang}, X.~W. and {Tang}, Z.~C. and {Tian}, J. and {Ting}, Samuel C.~C. and {Ting}, S.~M. and {Tomassetti}, N. and {Torsti}, J. and {T{\"u}ys{\"u}z}, C. and {Urban}, T. and {Usoskin}, I. and {Vagelli}, V. and {Vainio}, R. and {Valencia-Otero}, M. and {Valente}, E. and {Valtonen}, E. and {V{\'a}zquez Acosta}, M. and {Vecchi}, M.},
        title = "{Properties of Iron Primary Cosmic Rays: Results from the Alpha Magnetic Spectrometer}",
      journal = {prl},
         year = 2021,
        month = jan,
       volume = {126},
       number = {4},
          eid = {041104},
        pages = {041104},
          doi = {10.1103/PhysRevLett.126.041104},
       adsurl = {https://ui.adsabs.harvard.edu/abs/2021PhRvL.126d1104A}
}

@ARTICLE{2025ApJ...982L..33C,
       author = {{Cao}, Zhen and {Aharonian}, F. and {Bai}, Y.~X. and {Bao}, Y.~W. and {Bastieri}, D. and {Bi}, X.~J. and {Bi}, Y.~J. and {Bian}, W. and {Bukevich}, A.~V. and {Cai}, C.~M. and {Cao}, W.~Y. and {Cao}, Zhe and {Chang}, J. and {Chang}, J.~F. and {Chen}, A.~M. and {Chen}, E.~S. and {Chen}, H.~X. and {Chen}, Liang and {Chen}, Long and {Chen}, M.~J. and {Chen}, M.~L. and {Chen}, Q.~H. and {Chen}, S. and {Chen}, S.~H. and {Chen}, S.~Z. and {Chen}, T.~L. and {Chen}, X.~B. and {Chen}, X.~J. and {Chen}, Y. and {Cheng}, N. and {Cheng}, Y.~D. and {Chu}, M.~C. and {Cui}, M.~Y. and {Cui}, S.~W. and {Cui}, X.~H. and {Cui}, Y.~D. and {Dai}, B.~Z. and {Dai}, H.~L. and {Dai}, Z.~G. and {Danzengluobu} and {Diao}, Y.~X. and {Dong}, X.~Q. and {Duan}, K.~K. and {Fan}, J.~H. and {Fan}, Y.~Z. and {Fang}, J. and {Fang}, J.~H. and {Fang}, K. and {Feng}, C.~F. and {Feng}, H. and {Feng}, L. and {Feng}, S.~H. and {Feng}, X.~T. and {Feng}, Y. and {Feng}, Y.~L. and {Gabici}, S. and {Gao}, B. and {Gao}, C.~D. and {Gao}, Q. and {Gao}, W. and {Gao}, W.~K. and {Ge}, M.~M. and {Ge}, T.~T. and {Geng}, L.~S. and {Giacinti}, G. and {Gong}, G.~H. and {Gou}, Q.~B. and {Gu}, M.~H. and {Guo}, F.~L. and {Guo}, J. and {Guo}, X.~L. and {Guo}, Y.~Q. and {Guo}, Y.~Y. and {Han}, Y.~A. and {Hannuksela}, O.~A. and {Hasan}, M. and {He}, H.~H. and {He}, H.~N. and {He}, J.~Y. and {He}, X.~Y. and {He}, Y. and {Hernandez-Cadena}, S. and {Hor}, Y.~K. and {Hou}, B.~W. and {Hou}, C. and {Hou}, X. and {Hu}, H.~B. and {Hu}, S.~C. and {Huang}, C. and {Huang}, D.~H. and {Huang}, J.~J. and {Huang}, T.~Q. and {Huang}, W.~J. and {Huang}, X.~T. and {Huang}, X.~Y. and {Huang}, Y. and {Huang}, Y.~Y. and {Ji}, X.~L. and {Jia}, H.~Y. and {Jia}, K. and {Jiang}, H.~B. and {Jiang}, K. and {Jiang}, X.~W. and {Jiang}, Z.~J. and {Jin}, M. and {Kaci}, S. and {Kang}, M.~M. and {Karpikov}, I. and {Khangulyan}, D. and {Kuleshov}, D. and {Kurinov}, K. and {Li}, B.~B. and {Li}, Cheng and {Li}, Cong and {Li}, D. and {Li}, F. and {Li}, H.~B. and {Li}, H.~C. and {Li}, Jian and {Li}, Jie and {Li}, K. and {Li}, L. and {Li}, R.~L. and {Li}, S.~D. and {Li}, T.~Y. and {Li}, W.~L. and {Li}, X.~R. and {Li}, Xin and {Li}, Y.~Z. and {Li}, Zhe and {Li}, Zhuo and {Liang}, E.~W. and {Liang}, Y.~F. and {Lin}, S.~J. and {Liu}, B. and {Liu}, C. and {Liu}, D. and {Liu}, D.~B. and {Liu}, H. and {Liu}, H.~D. and {Liu}, J. and {Liu}, J.~L. and {Liu}, J.~R. and {Liu}, M.~Y. and {Liu}, R.~Y. and {Liu}, S.~M. and {Liu}, W. and {Liu}, X. and {Liu}, Y. and {Liu}, Y.~N. and {Lou}, Y.~Q. and {Luo}, Q. and {Luo}, Y. and {Lv}, H.~K. and {Ma}, B.~Q. and {Ma}, L.~L. and {Ma}, X.~H. and {Mao}, J.~R. and {Min}, Z. and {Mitthumsiri}, W. and {Mou}, G.~B. and {Mu}, H.~J. and {Nan}, Y.~C. and {Neronov}, A. and {Ng}, K.~C.~Y. and {Ni}, M.~Y. and {Nie}, L. and {Ou}, L.~J. and {Pattarakijwanich}, P. and {Pei}, Z.~Y. and {Qi}, J.~C. and {Qi}, M.~Y. and {Qin}, J.~J. and {Raza}, A. and {Ren}, C.~Y. and {Ruffolo}, D. and {S{\'a}iz}, A. and {Saeed}, M. and {Semikoz}, D. and {Shao}, L. and {Shchegolev}, O. and {Shen}, Y.~Z. and {Sheng}, X.~D. and {Shi}, Z.~D. and {Shu}, F.~W. and {Song}, H.~C. and {Stenkin}, Yu. V. and {Stepanov}, V. and {Su}, Y. and {Sun}, D.~X. and {Sun}, H. and {Sun}, Q.~N. and {Sun}, X.~N. and {Sun}, Z.~B. and {Tabasam}, N.~H. and {Takata}, J. and {Tam}, P.~H.~T. and {Tan}, H.~B. and {Tang}, Q.~W. and {Tang}, R.},
        title = "{Broadband {\ensuremath{\gamma}}-Ray Spectrum of Supernova Remnant Cassiopeia A}",
      journal = {apjl},
         year = 2025,
        month = mar,
       volume = {982},
       number = {1},
          eid = {L33},
        pages = {L33},
          doi = {10.3847/2041-8213/adb97c},
archivePrefix = {arXiv},
       eprint = {2502.04848},
 primaryClass = {astro-ph.HE},
       adsurl = {https://ui.adsabs.harvard.edu/abs/2025ApJ...982L..33C}
}

@ARTICLE{2023Univ....9...98S,
       author = {{Sinitsyna}, Vera G. and {Sinitsyna}, Vera Y.},
        title = "{Cosmic-Ray Acceleration in Supernova Remnants}",
      journal = {Universe},
         year = 2023,
        month = feb,
       volume = {9},
       number = {2},
          eid = {98},
        pages = {98},
          doi = {10.3390/universe9020098},
       adsurl = {https://ui.adsabs.harvard.edu/abs/2023Univ....9...98S}
}

@ARTICLE{2021MNRAS.502..472A,
       author = {{Araya}, M. and {Herrera}, C.},
        title = "{A source of gamma rays coincident with the shell of the supernova remnant CTB 80}",
      journal = {mnras},
         year = 2021,
        month = mar,
       volume = {502},
       number = {1},
        pages = {472-477},
          doi = {10.1093/mnras/stab101},
archivePrefix = {arXiv},
       eprint = {2101.04020},
 primaryClass = {astro-ph.HE},
       adsurl = {https://ui.adsabs.harvard.edu/abs/2021MNRAS.502..472A}
}

@ARTICLE{2025ApJ...988...93N,
       author = {{Nie}, Lin and {Guo}, Yi-Qing and {Liu}, Si-Ming},
        title = "{A Possible Common Physics Picture Reflected by the Gamma-Ray Emission of the Galactic Center}",
      journal = {apj},
         year = 2025,
        month = jul,
       volume = {988},
       number = {1},
          eid = {93},
        pages = {93},
          doi = {10.3847/1538-4357/ade3ce},
archivePrefix = {arXiv},
       eprint = {2503.05261},
 primaryClass = {astro-ph.HE},
       adsurl = {https://ui.adsabs.harvard.edu/abs/2025ApJ...988...93N}
}

@ARTICLE{2018PhRvD..97f3008G,
       author = {{Guo}, Yi-Qing and {Yuan}, Qiang},
        title = "{Understanding the spectral hardenings and radial distribution of Galactic cosmic rays and Fermi diffuse {\ensuremath{\gamma}} rays with spatially-dependent propagation}",
      journal = {prd},
         year = 2018,
        month = mar,
       volume = {97},
       number = {6},
          eid = {063008},
        pages = {063008},
          doi = {10.1103/PhysRevD.97.063008},
archivePrefix = {arXiv},
       eprint = {1801.05904},
 primaryClass = {astro-ph.HE},
       adsurl = {https://ui.adsabs.harvard.edu/abs/2018PhRvD..97f3008G}
}

@ARTICLE{2024PhRvD.109f3001Y,
       author = {{Yao}, Yu-Hua and {Dong}, Xu-Lin and {Guo}, Yi-Qing and {Yuan}, Qiang},
        title = "{Common origin of the multimessenger spectral anomaly of Galactic cosmic rays}",
      journal = {prd},
         year = 2024,
        month = mar,
       volume = {109},
       number = {6},
          eid = {063001},
        pages = {063001},
          doi = {10.1103/PhysRevD.109.063001},
archivePrefix = {arXiv},
       eprint = {2308.15866},
 primaryClass = {astro-ph.HE},
       adsurl = {https://ui.adsabs.harvard.edu/abs/2024PhRvD.109f3001Y}
}

@ARTICLE{2024ApJ...974..276N,
       author = {{Nie}, Lin and {Qian}, Xiang-Li and {Guo}, Yi-Qing and {Liu}, Si-Ming},
        title = "{Contribution of the Cygnus Bubble to the Galactic Cosmic Ray Spectrum and Diffuse {\ensuremath{\gamma}}-Ray Emissions}",
      journal = {apj},
         year = 2024,
        month = oct,
       volume = {974},
       number = {2},
          eid = {276},
        pages = {276},
          doi = {10.3847/1538-4357/ad7eab},
archivePrefix = {arXiv},
       eprint = {2408.01693},
 primaryClass = {astro-ph.HE},
       adsurl = {https://ui.adsabs.harvard.edu/abs/2024ApJ...974..276N}
}

@ARTICLE{2023ApJ...956...75Q,
       author = {{Qiao}, Bing-Qiang and {Guo}, Yi-Qing and {Liu}, Wei and {Bi}, Xiao-Jun},
        title = "{Nearby SNR: A Possible Common Origin of Multi-messenger Anomalies in the Spectra, Ratios, and Anisotropy of Cosmic Rays}",
      journal = {apj},
         year = 2023,
        month = oct,
       volume = {956},
       number = {2},
          eid = {75},
        pages = {75},
          doi = {10.3847/1538-4357/acf453},
archivePrefix = {arXiv},
       eprint = {2212.05641},
 primaryClass = {astro-ph.HE},
       adsurl = {https://ui.adsabs.harvard.edu/abs/2023ApJ...956...75Q}
}

@ARTICLE{2026JCAP...10..022N,
       author = {{Nie}, Lin and {Guo}, Yi-Qing and {Liu}, Si-Ming},
        title = "{Two-component {\ensuremath{\gamma}}-ray structure from the CR sources within dense clouds}",
      journal = {jcap},
         year = 2026,
        month = oct,
       volume = {2026},
       number = {10},
          eid = {022},
        pages = {022},
          doi = {10.1088/1475-7516/2026/10/022},
archivePrefix = {arXiv},
       eprint = {2602.07610},
 primaryClass = {astro-ph.HE},
       adsurl = {https://ui.adsabs.harvard.edu/abs/2026JCAP...10..022N}
}

@ARTICLE{1998ApJ...509..212S,
       author = {{Strong}, Andrew W. and {Moskalenko}, Igor V.},
        title = "{Propagation of Cosmic-Ray Nucleons in the Galaxy}",
      journal = {apj},
         year = 1998,
        month = dec,
       volume = {509},
       number = {1},
        pages = {212-228},
          doi = {10.1086/306470},
archivePrefix = {arXiv},
       eprint = {astro-ph/9807150},
 primaryClass = {astro-ph},
       adsurl = {https://ui.adsabs.harvard.edu/abs/1998ApJ...509..212S}
}

@ARTICLE{1978ApJ...221L..29B,
       author = {{Blandford}, R.~D. and {Ostriker}, J.~P.},
        title = "{Particle acceleration by astrophysical shocks.}",
      journal = {apjl},
         year = 1978,
        month = apr,
       volume = {221},
        pages = {L29-L32},
          doi = {10.1086/182658},
       adsurl = {https://ui.adsabs.harvard.edu/abs/1978ApJ...221L..29B}
}

@ARTICLE{2019MNRAS.482.3843T,
       author = {{Tang}, Xiaping},
        title = "{Gamma-ray emission from middle-aged supernova remnants interacting with molecular clouds: the challenge for current models}",
      journal = {mnras},
         year = 2019,
        month = jan,
       volume = {482},
       number = {3},
        pages = {3843-3856},
          doi = {10.1093/mnras/sty1042},
archivePrefix = {arXiv},
       eprint = {1707.00958},
 primaryClass = {astro-ph.HE},
       adsurl = {https://ui.adsabs.harvard.edu/abs/2019MNRAS.482.3843T}
}

@ARTICLE{2011NatCo...2..194M,
       author = {{Malkov}, M.~A. and {Diamond}, P.~H. and {Sagdeev}, R.~Z.},
        title = "{Mechanism for spectral break in cosmic ray proton spectrum of supernova remnant W44}",
      journal = {Nature Communications},
         year = 2011,
        month = feb,
       volume = {2},
          eid = {194},
        pages = {194},
          doi = {10.1038/ncomms1195},
archivePrefix = {arXiv},
       eprint = {1004.4714},
 primaryClass = {astro-ph.HE},
       adsurl = {https://ui.adsabs.harvard.edu/abs/2011NatCo...2..194M}
}

@ARTICLE{1982ApJ...260..625B,
       author = {{Blandford}, R.~D. and {Cowie}, L.~L.},
        title = "{Radio emission from supernova remnants in a cloudy interstellar medium}",
      journal = {apj},
         year = 1982,
        month = sep,
       volume = {260},
        pages = {625-634},
          doi = {10.1086/160284},
       adsurl = {https://ui.adsabs.harvard.edu/abs/1982ApJ...260..625B}
}

@INPROCEEDINGS{2015ICRC...34..922Z,
       author = {{Zabalza}, V.},
        title = "{Naima: a Python package for inference of particle distribution properties from nonthermal spectra}",
    booktitle = {34th International Cosmic Ray Conference (ICRC2015)},
         year = 2015,
       series = {International Cosmic Ray Conference},
       volume = {34},
        month = jul,
          eid = {922},
        pages = {922},
          doi = {10.22323/1.236.0922},
archivePrefix = {arXiv},
       eprint = {1509.03319},
 primaryClass = {astro-ph.HE},
       adsurl = {https://ui.adsabs.harvard.edu/abs/2015ICRC...34..922Z}
}

@ARTICLE{2006PhRvC..74c4318O,
       author = {{Orce}, J.~N. and {Bruce}, A.~M. and {Emmanouilidis}, A. and {Byrne}, A.~P. and {Dracoulis}, G.~D. and {Kib{\'e}di}, T. and {Caama{\~n}o}, M. and {El-Masri}, H. and {Pearson}, C.~J. and {Podoly{\'a}k}, Zs. and {Stevenson}, P.~D. and {Walker}, P.~M. and {Xu}, F.~R. and {Cullen}, D.~M. and {Wheldon}, C.},
        title = "{Shape-driving effects in the triaxial nucleus, Xe128}",
      journal = {prc},
         year = 2006,
        month = sep,
       volume = {74},
       number = {3},
          eid = {034318},
        pages = {034318},
          doi = {10.1103/PhysRevC.74.034318},
       adsurl = {https://ui.adsabs.harvard.edu/abs/2006PhRvC..74c4318O}
}

@ARTICLE{2013PASP..125..306F,
       author = {{Foreman-Mackey}, Daniel and {Hogg}, David W. and {Lang}, Dustin and {Goodman}, Jonathan},
        title = "{emcee: The MCMC Hammer}",
      journal = {pasp},
         year = 2013,
        month = mar,
       volume = {125},
       number = {925},
        pages = {306},
          doi = {10.1086/670067},
archivePrefix = {arXiv},
       eprint = {1202.3665},
 primaryClass = {astro-ph.IM},
       adsurl = {https://ui.adsabs.harvard.edu/abs/2013PASP..125..306F}
}

@ARTICLE{2018ApJS..238...35S,
       author = {{Shan}, S.~S. and {Zhu}, H. and {Tian}, W.~W. and {Zhang}, M.~F. and {Zhang}, H.~Y. and {Wu}, D. and {Yang}, A.~Y.},
        title = "{Distances of Galactic Supernova Remnants Using Red Clump Stars}",
      journal = {apjs},
         year = 2018,
        month = oct,
       volume = {238},
       number = {2},
          eid = {35},
        pages = {35},
          doi = {10.3847/1538-4365/aae07a},
archivePrefix = {arXiv},
       eprint = {1810.06014},
 primaryClass = {astro-ph.IM},
       adsurl = {https://ui.adsabs.harvard.edu/abs/2018ApJS..238...35S}
}

@ARTICLE{2012MNRAS.423..718L,
       author = {{Leahy}, D.~A. and {Ranasinghe}, S.},
        title = "{Radio observations of CTB80: detection of the snowplough in an old supernova remnant}",
      journal = {mnras},
         year = 2012,
        month = jun,
       volume = {423},
       number = {1},
        pages = {718-724},
          doi = {10.1111/j.1365-2966.2012.20909.x},
       adsurl = {https://ui.adsabs.harvard.edu/abs/2012MNRAS.423..718L}
}

@ARTICLE{1990ApJ...364..178K,
       author = {{Koo}, Bon-Chul and {Reach}, William T. and {Heiles}, Carl and {Fesen}, Robert A. and {Shull}, J. Michael},
        title = "{Detection of an Expanding H i Shell in the Old Supernova Remnant CTB 80}",
      journal = {apj},
         year = 1990,
        month = nov,
       volume = {364},
        pages = {178},
          doi = {10.1086/169400},
       adsurl = {https://ui.adsabs.harvard.edu/abs/1990ApJ...364..178K}
}

@ARTICLE{1993ApJ...417..196K,
       author = {{Koo}, Bon-Chul and {Yun}, Min-Su and {Ho}, Paul T.~P. and {Lee}, Youngung},
        title = "{Interaction between the Supernova Remnant CTB 80 and the Ambient Interstellar Medium: H i and CO Observations}",
      journal = {apj},
         year = 1993,
        month = nov,
       volume = {417},
        pages = {196},
          doi = {10.1086/173303},
       adsurl = {https://ui.adsabs.harvard.edu/abs/1993ApJ...417..196K}
}

@ARTICLE{2024ApJS..271...25C,
       author = {{Cao}, Zhen and {Aharonian}, F. and {An}, Q. and {Axikegu} and {Bai}, Y.~X. and {Bao}, Y.~W. and {Bastieri}, D. and {Bi}, X.~J. and {Bi}, Y.~J. and {Cai}, J.~T. and {Cao}, Q. and {Cao}, W.~Y. and {Cao}, Zhe and {Chang}, J. and {Chang}, J.~F. and {Chen}, A.~M. and {Chen}, E.~S. and {Chen}, Liang and {Chen}, Lin and {Chen}, Long and {Chen}, M.~J. and {Chen}, M.~L. and {Chen}, Q.~H. and {Chen}, S.~H. and {Chen}, S.~Z. and {Chen}, T.~L. and {Chen}, Y. and {Cheng}, N. and {Cheng}, Y.~D. and {Cui}, M.~Y. and {Cui}, S.~W. and {Cui}, X.~H. and {Cui}, Y.~D. and {Dai}, B.~Z. and {Dai}, H.~L. and {Dai}, Z.~G. and {Danzengluobu} and {Della Volpe}, D. and {Dong}, X.~Q. and {Duan}, K.~K. and {Fan}, J.~H. and {Fan}, Y.~Z. and {Fang}, J. and {Fang}, K. and {Feng}, C.~F. and {Feng}, L. and {Feng}, S.~H. and {Feng}, X.~T. and {Feng}, Y.~L. and {Gabici}, S. and {Gao}, B. and {Gao}, C.~D. and {Gao}, L.~Q. and {Gao}, Q. and {Gao}, W. and {Gao}, W.~K. and {Ge}, M.~M. and {Geng}, L.~S. and {Giacinti}, G. and {Gong}, G.~H. and {Gou}, Q.~B. and {Gu}, M.~H. and {Guo}, F.~L. and {Guo}, X.~L. and {Guo}, Y.~Q. and {Guo}, Y.~Y. and {Han}, Y.~A. and {He}, H.~H. and {He}, H.~N. and {He}, J.~Y. and {He}, X.~B. and {He}, Y. and {Heller}, M. and {Hor}, Y.~K. and {Hou}, B.~W. and {Hou}, C. and {Hou}, X. and {Hu}, H.~B. and {Hu}, Q. and {Hu}, S.~C. and {Huang}, D.~H. and {Huang}, T.~Q. and {Huang}, W.~J. and {Huang}, X.~T. and {Huang}, X.~Y. and {Huang}, Y. and {Huang}, Z.~C. and {Ji}, X.~L. and {Jia}, H.~Y. and {Jia}, K. and {Jiang}, K. and {Jiang}, X.~W. and {Jiang}, Z.~J. and {Jin}, M. and {Kang}, M.~M. and {Ke}, T. and {Kuleshov}, D. and {Kurinov}, K. and {Li}, B.~B. and {Li}, Cheng and {Li}, Cong and {Li}, D. and {Li}, F. and {Li}, H.~B. and {Li}, H.~C. and {Li}, H.~Y. and {Li}, J. and {Li}, Jian and {Li}, Jie and {Li}, K. and {Li}, W.~L. and {Li}, W.~L. and {Li}, X.~R. and {Li}, Xin and {Li}, Y.~Z. and {Li}, Zhe and {Li}, Zhuo and {Liang}, E.~W. and {Liang}, Y.~F. and {Lin}, S.~J. and {Liu}, B. and {Liu}, C. and {Liu}, D. and {Liu}, H. and {Liu}, H.~D. and {Liu}, J. and {Liu}, J.~L. and {Liu}, J.~Y. and {Liu}, M.~Y. and {Liu}, R.~Y. and {Liu}, S.~M. and {Liu}, W. and {Liu}, Y. and {Liu}, Y.~N. and {Lu}, R. and {Luo}, Q. and {Lv}, H.~K. and {Ma}, B.~Q. and {Ma}, L.~L. and {Ma}, X.~H. and {Mao}, J.~R. and {Min}, Z. and {Mitthumsiri}, W. and {Mu}, H.~J. and {Nan}, Y.~C. and {Neronov}, A. and {Ou}, Z.~W. and {Pang}, B.~Y. and {Pattarakijwanich}, P. and {Pei}, Z.~Y. and {Qi}, M.~Y. and {Qi}, Y.~Q. and {Qiao}, B.~Q. and {Qin}, J.~J. and {Ruffolo}, D. and {S{\'a}iz}, A. and {Semikoz}, D. and {Shao}, C.~Y. and {Shao}, L. and {Shchegolev}, O. and {Sheng}, X.~D. and {Shu}, F.~W. and {Song}, H.~C. and {Stenkin}, Yu. V. and {Stepanov}, V. and {Su}, Y. and {Sun}, Q.~N. and {Sun}, X.~N. and {Sun}, Z.~B. and {Tam}, P.~H.~T. and {Tang}, Q.~W. and {Tang}, Z.~B. and {Tian}, W.~W. and {Wang}, C. and {Wang}, C.~B. and {Wang}, G.~W. and {Wang}, H.~G. and {Wang}, H.~H. and {Wang}, J.~C. and {Wang}, K. and {Wang}, L.~P. and {Wang}, L.~Y. and {Wang}, P.~H. and {Wang}, R. and {Wang}, W. and {Wang}, X.~G. and {Wang}, X.~Y. and {Wang}, Y. and {Wang}, Y.~D. and {Wang}, Y.~J. and {Wang}, Z.~H. and {Wang}, Z.~X. and {Wang}, Zhen and {Wang}, Zheng and {Wei}, D.~M. and {Wei}, J.~J. and {Wei}, Y.~J. and {Wen}, T. and {Wu}, C.~Y. and {Wu}, H.~R.},
        title = "{The First LHAASO Catalog of Gamma-Ray Sources}",
      journal = {apjs},
         year = 2024,
        month = mar,
       volume = {271},
       number = {1},
          eid = {25},
        pages = {25},
          doi = {10.3847/1538-4365/acfd29},
archivePrefix = {arXiv},
       eprint = {2305.17030},
 primaryClass = {astro-ph.HE},
       adsurl = {https://ui.adsabs.harvard.edu/abs/2024ApJS..271...25C}
}

%% or
%% [B] Manual formatting (see below)
%% (i) We suggest to always provide author, title and journal data or doi:
%% in short all the informations that clearly identify a document.
%% (ii) please avoid comments such as "For a review'', "For some examples",
%% "and references therein" or move them in the text. In general, please leave only references in the bibliography and move all
%% accessory text in footnotes.
%% (iii) Also, please have only one work for each \bibitem.

%\begin{thebibliography}{99}

%\bibitem{a}
%Author,
%\emph{Title},
%\emph{J. Abbrev.} {\bf vol} (year) pg.

%\bibitem{b}
%Author,
%\emph{Title},
%arxiv:1234.5678.

%\bibitem{c}

%\emph{Title},
%Publisher (year).

%\end{thebibliography}
\end{document}